\documentclass[journal]{IEEEtran}

\usepackage{xcolor,soul,framed} 

\colorlet{shadecolor}{yellow}
\usepackage[pdftex]{graphicx}
\graphicspath{{../pdf/}{../jpeg/}}
\DeclareGraphicsExtensions{.pdf,.jpeg,.png}

\usepackage[cmex10]{amsmath}
\usepackage{amssymb}
\let\labelindent\relax
\usepackage[inline,shortlabels]{enumitem}

\usepackage{mathabx}
\usepackage{array}
\usepackage{mdwmath}
\usepackage{mdwtab}
\usepackage{eqparbox}
\usepackage{url}
\usepackage{hyperref}
\usepackage[nameinlink]{cleveref} 
\usepackage{doi}  
\usepackage{cite}
\usepackage{adjustbox}
\usepackage{booktabs,multirow,makecell,threeparttable,siunitx}
\usepackage[skip=2pt]{caption,subcaption}
\newcommand{\ua}{\textsc{UA}}
\newcommand{\wa}{\textsc{WA}}
\newcommand{\mf}{\textsc{F1}}

\newcommand{\mathbold}[1]{\ensuremath{\boldsymbol{\mathbf{#1}}}}

\newcommand{\nestedmathbold}[1]{{\mathbold{#1}}}

\newcommand{\mbi}{\nestedmathbold{i}}
\newcommand{\mbj}{\nestedmathbold{j}}
\newcommand{\mbk}{\nestedmathbold{k}}

\newcommand{\mbm}{\nestedmathbold{m}}

\newcommand{\mbp}{\nestedmathbold{p}}
\newcommand{\mbq}{\nestedmathbold{q}}

\newcommand{\mbx}{\nestedmathbold{x}}
\newcommand{\mby}{\nestedmathbold{y}}
\newcommand{\mbz}{\nestedmathbold{z}}

\newcommand{\mbE}{\nestedmathbold{E}}

\newcommand{\mbG}{\nestedmathbold{G}}

\newcommand{\cY}{\mathcal{Y}}

\newcommand{\bbH}{\mathbb{H}}
\newcommand{\bbR}{\mathbb{R}}

\Crefname{figure}{Fig.}{Figs.}
\Crefname{equation}{Eq.}{Eqs.}
\Crefname{section}{Sec.}{Secs.}

\makeatletter

\newcommand{\footnoterecall}[1]{\textsuperscript{\hyperref[#1]{\ref*{#1}}}}
\makeatother

\begin{document}
\bstctlcite{IEEEexample:BSTcontrol}
    \title{Learning Physiology-Informed Vocal Spectrotemporal Representations for Speech Emotion Recognition}
\author{
Xu~Zhang,
Longbing~Cao\textsuperscript{*},~\IEEEmembership{Senior Member,~IEEE},
Zhangkai~Wu,
and~Runze~Yang%
\thanks{Xu Zhang, Longbing Cao, Zhangkai Wu, and Runze Yang are with the Frontier AI Research Centre, Macquarie University, Sydney, NSW, Australia
(e-mail: xu.zhang12@hdr.mq.edu.au; longbing.cao@mq.edu.au; zhangkai.wu@mq.edu.au; runze.yang@hdr.mq.edu.au).}%
\thanks{Runze Yang is also with Shanghai Jiao Tong University, Shanghai, China.}%
\thanks{\textsuperscript{*}Corresponding author.}%
}

\maketitle

\begin{abstract}
    Speech emotion recognition (SER) is essential for humanoid robot tasks such as social robotic interactions and robotic psychological diagnosis, where interpretable and efficient models are critical for safety and performance. Existing deep models trained on large datasets remain largely uninterpretable, often insufficiently modeling underlying emotional acoustic signals and failing to capture and analyze the core physiology of emotional vocal behaviors. Physiological research on human voices shows that the dynamics of vocal amplitude and phase correlate with emotions through the vocal tract filter and the glottal source. However, most existing deep models solely involve amplitude but fail to couple the physiological features of and between amplitude and phase. Here, we propose PhysioSER, a physiology-informed vocal spectrotemporal representation learning method, to address these issues by a compact, plug-and-play design. PhysioSER constructs amplitude–phase views informed by voice anatomy and physiology (VAP) to complement SSL models for SER. This VAP-informed framework incorporates two parallel workflows: a vocal feature representation branch to decompose vocal signals based on VAP, embed them into a quaternion field, and use Hamilton structured quaternion convolutions for modeling their dynamic interactions; and a latent representation branch based on a frozen SSL backbone. Then, utterance-level features from both workflows are aligned by a Contrastive Projection and Alignment framework, followed by a shallow attention fusion head for SER classification. PhysioSER is shown interpretable and efficient for SER through extensive evaluations across 14 datasets, 10 languages, and 6 backbones, and its practical efficacy is validated by real-time deployment on a humanoid robotic platform.
\end{abstract} 

 
\IEEEpeerreviewmaketitle
\section{Introduction}
    \IEEEPARstart{S}{peech} emotion recognition (SER) is essential for tasks including human-humanoid interaction, social humanoid robots, and intelligent applications such as assisting clinical assessment \cite{sheng2024deep, guo2025enhancing, cao2025humanoid} and for developing emotional humanoids \cite{cao2025humanoid,Cao2025HumanoidAI}. Recognizing emotions in speech has migrated from handcrafted prosodic and spectral descriptors \cite{10.1016/j.patcog.2010.09.020, anagnostopoulos2015features} to self-supervised learning (SSL) encoders that learn transferable acoustic representations from massive unlabeled corpora, achieving state-of-the-art performance on SER benchmarks \cite{baevski2020wav2vec, chen2022wavlm, 10.1109/TASLP.2021.3122291, ma-etal-2024-emotion2vec, 10.5555/3618408.3618611, elizalde2023clap}.

    Despite great success, these data-driven modeling methods lack explicit interpretability and often rely on large datasets or coarse speech cues, overlooking signal dynamics driven by speech production. This limitation reflects a disconnection from voice production mechanisms, where affect jointly modulates the glottal source and the vocal tract filter~\cite{quatieri2002shape, deshmukh2005use, hegde2006significance}. Widely used approaches based on spectrogram representations rely on Mel and related magnitude features, and many disentanglement methods still build on these magnitude representations~\cite{ancilin2021improved, xie2025survey, elizalde2023clap, 10.5555/3618408.3618611}, thereby underusing phase information~\cite{qian2020unsupervised, williamson2015complex, erdogan2015phase}. Large scale waveform SSL models typically optimize masked prediction or denoising objectives without explicitly reconstructing or regularizing short-time Fourier transform (STFT) phase~\cite{baevski2020wav2vec, 10.1109/TASLP.2021.3122291, chen2022wavlm, ma-etal-2024-emotion2vec}. On the other, many methods incorporate both magnitude and phase, while they often either concatenate them as separate features or treat them as independent channels~\cite{hegde2006significance, vijayan2015analysis}, rather than modeling the physiology-informed couplings between them. As a result, the structured interdependencies dictated by voice production mechanisms remain insufficiently captured, potentially weakening cross component interactions. Therefore, the central challenge is not to simply include these signals, but to develop a principled framework that explicitly models their coupling dynamics by following their principles such as the physiology of emotional speech production.
    
    In speech production physiology, studies show that emotion modulates the glottal source and speech energy and rhythm through systematic changes in respiration and subglottal pressure, propagating along the production chain from respiration, through the glottis, to the vocal tract \cite{kreibig2010autonomic, scherer2003vocal}. These modulations interact with vocal tract resonances, yielding nontrivial covariation between the glottal source and the vocal tract filter, and leaving measurable acoustic correlates in speech \cite{titze2008nonlinear, sundberg1993short}. This physiology informed production process can be characterized by four complementary emotional related features: the log-spectral magnitude ($M$), providing cues about vocal tract configuration \cite{story2001relationship}; its time derivative ($\rho$), capturing energy flow and fluctuation rhythms often associated with respiratory support and stress patterns \cite{bello2005tutorial}; the instantaneous frequency ($f_{\text{inst}}$), reflecting the glottal oscillation rate and carrying fundamental frequency and microprosody \cite{boashash2002estimating}; and the group delay ($\tau_g$), which is informative of vocal tract resonance and glottal pulse details \cite{yegnanarayana2022group}. Crucially, these features are synergistic products of a unified physiological mechanism. Treating amplitude and phase features in isolation serves the inherent linkage, which is jointly governed by voice production mechanisms and acoustic physics \cite{titze2008nonlinear, yegnanarayana2022group, hegde2006significance}. This necessitates a physiology informed, structurally coupled representation that employs constrained cross-channel mixing to explicitly encode coupled dynamics of amplitude and phase features. We find that operations satisfying Hamilton coupling symmetry provide a natural inductive bias for this purpose. As we detail in \Cref{analyze_signals} from physiological, physical, and mathematical perspectives, this structure efficiently extracts key discriminative information about emotional variation while preserving this intrinsic signal coupling.

    Therefore, we propose \textbf{Physio}logy-Informed Vocal Spectrotemporal Representation Learning for \textbf{S}peech \textbf{E}motion \textbf{R}ecognition (PhysioSER), a framework parallel representation workflows  that integrates a general latent representation workflow with a physiology informed vocal representation workflow for SER. PhysioSER comprises four main components: (i) a latent speech representation workflow to extract general-purpose features from raw speech using a frozen SSL backbone; (ii) a voice anatomy and physiology (VAP)-informed representation workflow to decompose emotional speech into a compact, physically and physiologically consistent quartet of signals derived from Fourier analysis; (iii) a Hamilton structured Quaternion Spectrotemporal Encoder (QSE) to embed the quartet (${M,\rho,f_{\mathrm{inst}},\tau_g}$) as a structured four channels representation and to mix them via quaternion convolutions, providing cross-channel coupling to capture amplitude-phase dynamics with structured parameter sharing~\cite{zhu2018quaternion}; and (iv) a lightweight Contrastive Projection and Alignment (CPA) framework that aligns utterance-level latent and vocal features to exploit their complementarity, followed by a shallow attention-based fusion head for classification. Overall, PhysioSER retains the latent feature from SSL while injecting VAP-informed inductive biases to improve interpretability and capture production grounded cues. It consistently improves weighted accuracy (WA), unweighted accuracy (UA), and macro-F1 over the corresponding backbones. On dataset CREMA-D with a frozen WavLM \cite{chen2022wavlm}, WA increases from 69.69\% to 75.20\%. By training only $2\%$ of backbone parameters, PhysioSER surpasses a fully fine-tuned WavLM. Gains are larger for weaker backbones and for non-English datasets, suggesting PhysioSER captures robust and language agnostic discriminative information. Our contributions include:
    \begin{itemize}[noitemsep, topsep=0pt]
    \item decomposing speech from a voice anatomy and physiology (VAP)-informed perspective and formalizing a compact yet discriminative quartet to comprehensively capture emotional acoustic cues.
    \item the Quaternion Spectrotemporal Encoder (QSE) to mix the quartet via Hamilton structured quaternion convolutions for modeling coupled glottal source and vocal tract dynamics with high parameter efficiency, providing cross-channel coupling.
    \item a Contrastive Projection and Alignment (CPA) framework to align the utterance-level vocal and latent representation spaces, with a lightweight attention-based fusion head to use their complementarity.
    \end{itemize}

    Across 14 datasets and 6 backbones, PhysioSER consistently improves accuracy, generalizes well across languages, and improves efficiency. We also validate each module via ablation studies and show the effect on humanoid Ameca enabling vocal emotions.
    The remainder of this paper is organized as follows: \Cref{sec:related_work} reviews the main developments in SER; \Cref{sec:preliminaries} introduces the necessary preliminaries and notation; \Cref{sec:method} presents the proposed PhysioSER framework and its modules; \Cref{sec:experiments} describes the comprehensive experimental results and the deployment on the humanoid robot; and \Cref{sec:conclusion} concludes the paper and discusses limitations and future directions.

\section{Related Work}
\label{sec:related_work}
    Early work in SER relied on handcrafted acoustic features grounded in theories of speech production and human perception \cite{xie2025survey}. The community commonly distinguished prosodic features such as fundamental frequency, energy, and duration, which capture suprasegmental structure, from spectral features such as MFCCs, which characterize the vocal tract filter \cite{9543566, bitouk2010class}. Standardized toolkits and feature sets, most notably openSMILE \cite{eyben2010opensmile} and GeMAPS \cite{DBLP:journals/taffco/EybenSSSABDELNT16}, systematized this paradigm and enabled reproducible benchmarks across the field. Despite their impact, fixed statistical aggregation limits the ability to capture temporal organization and locally salient patterns of emotional expression.

    Consequently, the field shifted toward automatic representation learning. Many researchers applied CNNs directly to 2D spectrograms, casting SER as an image classification task \cite{mao2014learning,satt2017efficient}. This approach was soon augmented with RNNs and LSTMs to model temporal dependencies explicitly \cite{xie2019speech,lee2015high-level}. Hybrid models that combine CNNs and LSTMs, often with attention mechanisms, became mainstream, enabling models to weight emotionally salient frames and learn more robust utterance level representations \cite{mirsamadi2017automatic,tzirakis2018end}. Despite their success, these supervised models require large amounts of labeled data, which are costly to acquire and inherently subjective.

    Self-supervised learning (SSL) mitigates label scarcity through a pre-training and fine-tuning paradigm \cite{gui2024survey}. The Wav2vec2 framework introduces masked contrastive learning with product quantization \cite{baevski2020wav2vec}. HuBERT replaces contrastive objectives with masked prediction of discrete hidden units discovered by offline clustering \cite{10.1109/TASLP.2021.3122291}. WavLM advances this line by adding denoising and utterance mixing so that the representation retains information about content, speaker identity, and paralinguistic features \cite{chen2022wavlm}. Subsequent work diverges by domain goals, including Emotion2Vec for affect-oriented pre-training \cite{ma-etal-2024-emotion2vec}, BEATs for general acoustic scenes \cite{10.5555/3618408.3618611}, and CLAP for alignment between audio and natural language \cite{elizalde2023clap}. Nevertheless, most SSL pipelines remain magnitude centric, meaning that representations are driven mainly by spectral envelopes and energy, while the phase spectrum is typically ignored \cite{yegnanarayana2022analysis}. This omission prevents accurate modeling of the physics of speech production, because emotional state affects both the vocal tract filter and the glottal source, and phase derivatives provide informative cues about vocal dynamics.

    Recent SER studies increasingly combine amplitude and phase. These descriptors are commonly aggregated with comprehensive statistical functionals to form vectors of fixed length, which are then classified using support vector machines (SVMs) or related models. Simple fusion pairs magnitude with phase derivative features in CNN or SVM pipelines \cite{guo2018speech, guo2022drp,shankar2024amp_phase}. These methods yield consistent gains but rely on hand-crafted descriptors and late concatenation, which leaves amplitude and phase loosely coupled and risks entangling affect with nuisance factors. Learned interaction models such as gated amplitude-phase transformers (APIN) and parallel magnitude-phase autoencoders \cite{guo2025apin, gudmalwar2022magnitude} move beyond static fusion, yet they lack a physiologically grounded analysis and do not explicitly encode the coupling between glottal source dynamics and vocal tract filtering in the network structure. Correlation-driven approaches align MFCC (amplitude) with MODGD (phase) using model DCCA before sequence modeling \cite{dcca2023}, but correlation maximization without production priors is vulnerable to nuisance correlations and offers limited interpretability. Implicit schemes like complex cepstra bundle amplitude and phase into a single representation \cite{patnaik2023speech}, improving accuracy while obscuring which phase mechanisms contribute. Ensemble designs that stack magnitude with MGDCC or DRP further raise accuracy \cite{shankar2024amp_phase}, at the cost of heavy computation and brittle fusion rules.

    As discussed, existing methods either overlook phase entirely or fuse amplitude and phase in a loosely coupled manner, making them vulnerable to nuisance correlations and lacking clear physiological grounding. In contrast, we explicitly model this joint structure by using a VAP-informed analysis of emotional speech production to motivate a structured amplitude–phase representation and coupling mechanism, and we fuse it with the latent representation from a pre-trained SSL backbone to provide a complementary feature set for a principled and interpretable SER framework.

    In addition, SER plays an increasingly essential role for humanlike robots especially humanoid robots. Humanoids are increasingly applied for social and emotional applications and activities, such as aged and elderly care, while relevant SER for emotional humanoids is very limited \cite{cao2025humanoid}. In this work, our physiology-informed learning of vocal spectrotemporal representations not only address the above SER gaps but also create potential for emotional humanoids. 

\section{Preliminaries}
\label{sec:preliminaries}
    In this section, we formalize the problem and establish notations for the variable length sequences used in our method. We then provide a brief overview of the quaternion algebra.
    
\subsection{Problem Formulation and Notations}
\label{problem}
    We formulate SER as a multi-class classification task. Let $\mbx$ denote the raw speech utterance and $\cY=\{1,\dots,C\}$ the emotion label set.
    From $\mbx$ we first construct a frame level acoustic sequence by segmenting the waveform with a Hamming window. This yields a variable length feature sequence $\mathcal{X}^\ast \in \bbR^{T^\ast \times F}$, where $T^\ast$ denotes the number of valid frames and $F$ the per frame feature dimension. We then pad each sequence with zeros to the batch maximum length $T$, obtaining the final model input $\mathcal{X} \in \bbR^{T \times F}$. The goal is to learn a mapping $f_{\Theta}: \bbR^{T \times F} \to \bbR^{C}$. The model, parameterized by $\Theta$, produces logits $\mbz=f_{\Theta}(\mathcal{X})\in\bbR^{C}$ and class probabilities $\mbp=\mathrm{Softmax}(\mbz)\in\Delta^{C-1}$. The final emotion class is then determined by $\hat{y} = \arg\max_{l\in\{1,\dots,C\}} (\mbp)_l$.

    To prevent padded positions from affecting computation, we use a binary mask $\mbm\in\{0,1\}^{T}$ with entries:
\begin{equation}
    (\mbm)_t =
    \begin{cases}
        1, & \text{if } t \le T^\ast,\\
        0, & \text{otherwise}.
    \end{cases}
\end{equation}
    Unless specified, all frame level operations (e.g., pooling and attention) are mask-aware and ignore padded positions.

\subsection{Quaternion Algebra Fundamentals}
    We summarize the quaternion algebra used in our method. A quaternion $q\in\bbH$ is a hypercomplex division algebra extending the complex numbers, defined as
\begin{equation}
    q \;=\; r + a\,\mbi + b\,\mbj + c\,\mbk, \qquad r,a,b,c\in\bbR,
\label{Qalgebra}
\end{equation}
    where $r\in\bbR$ is the real component and $a,b,c\in\bbR$ are the coefficients of the imaginary units $\mbi,\mbj,\mbk$, which satisfy:
\begin{equation}
    \mbi^2=\mbj^2=\mbk^2=\mbi\mbj\mbk=-1,\qquad 
    \mbi\mbj=\mbk,\ \mbj\mbi=-\mbk.
\end{equation}
    
    We represent $q$ as the real 4 dimensional vector $[r,a,b,c]^\top$ and view quaternion valued sequences as 4 real-valued channels stacked along the feature axis. Quaternion algebra imposes a structured coupling of parameters across these 4 components, which allows quaternion computations to be implemented with standard real-valued operators.

\section{Methodology}
\label{sec:method}
\begin{figure*}
  \begin{center}
  \includegraphics[width=\textwidth]{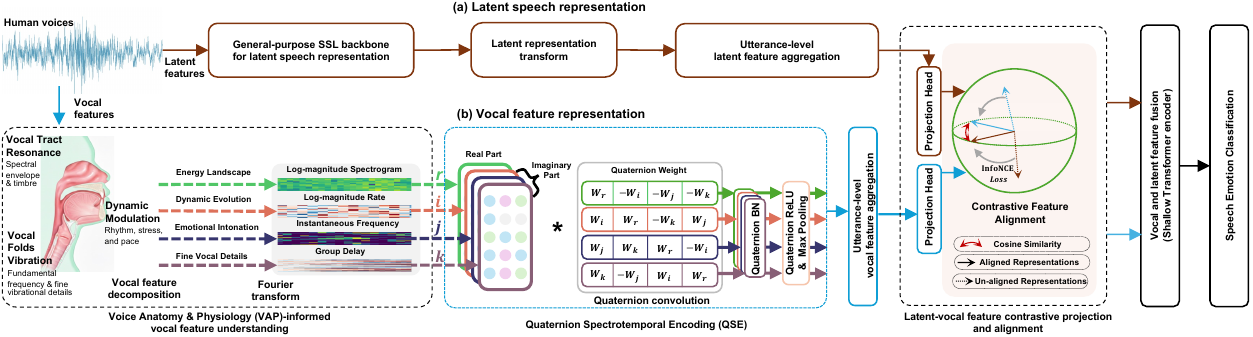}
  \caption{
  Structure of the PhysioSER. The model consists of two parallel workflows: (a) an upper latent speech representation workflow to extract general features from the raw waveform using a frozen SSL backbone, followed by a latent representation transform; and (b) a lower vocal feature representation workflow to decompose vocal signals based on voice anatomy and physiology (VAP)-informed knowledge. In (b), the Hamilton structured Quaternion Spectrotemporal Encoder (QSE) embeds a physiology-aligned quartet—log-magnitude ($M$), log-magnitude rate ($\rho$) (spectral flux), instantaneous frequency ($f_{\mathrm{inst}}$), and group delay ($\tau_g$)—and models their structured, dynamic interactions. The two branches are separately summarized into utterance-level embeddings and aligned by the Contrastive Projection and Alignment (CPA) framework via projection heads with an InfoNCE objective. Finally, a shallow Transformer encoder fuses the aligned latent and vocal representations for SER.
}
  \label{structure}
  \end{center}
\end{figure*}

    As shown in \Cref{structure}, PhysioSER enhances SER by integrating (i) latent speech representations extracted by a frozen, general-purpose SSL backbone, and (ii) VAP-informed vocal representations derived from an explicit amplitude-phase decomposition. Specifically, the vocal workflow performs a Fourier-based analysis to obtain a physiology informed quartet, which is encoded by a Hamilton structured QSE to capture structured cross-component interactions. Both workflows are summarized into utterance-level embeddings, aligned via projection heads using the CPA objective with an InfoNCE loss, and finally fused by a shallow Transformer encoder for emotion classification. We next detail the vocal feature representation workflow (\Cref{structure}b).    
    
\subsection{VAP-Informed Vocal Feature Decomposition}
\label{analyze_signals}
    We extract emotion related vocal representation from physiological and physical principles to construct an interpretable and VAP-informed representation.
    
\subsubsection{Physiological \& Physical Foundations}
    Speech emotion originates from physiological changes in vocal production \cite{kreibig2010autonomic, sauter2010cross}. The primary mechanism is the coupling between the glottal source and the vocal tract filter, whereby changes in laryngeal tension and respiration interact with vocal tract configuration \cite{li2018contributions, zhang2023influence}. Therefore, a comprehensive set of vocal features for emotion should capture the distinct contributions of the vocal source, the filter, and their dynamic interaction. Based on this insight, we extract core factors of speech emotion that align with three fundamental physiological dimensions of the vocal production system.

    \textbf{Vocal fold} vibration constitutes the phonatory source and is modulated by emotional state via laryngeal muscle control, which governs the vibratory frequency and pattern of the vocal folds, thereby shaping the emotional intonation of an utterance \cite{zhang2023influence}. To capture these source dynamics, we use 2 key speech representations. First, the \textbf{Instantaneous Frequency} (IF) provides a fine-grained proxy for the vocal fold oscillation rate and captures the fundamental intonation contour, a primary carrier of emotional prosody \cite{boashash2002estimating}. Second, we model fine vocal details related to the glottal pulse shape. For instance, heightened emotional tension increases vocal fold adduction, resulting in a steeper glottal pulse waveform \cite{baker2023high}. This change systematically alters phase relationships among frequency components propagating through the vocal tract, which are effectively captured by the \textbf{Group Delay} \cite{hegde2006significance, yegnanarayana2022group}.

    The \textbf{vocal tract} functions as a tunable resonant cavity whose structure is reconfigured by articulatory settings such as tongue position, lip rounding, and epilaryngeal narrowing \cite{story2001relationship}. These configurations shape the spectral envelope, which underlies the perceptual quality of timbre, the attribute that distinguishes sounds of identical pitch and loudness \cite{fabiani2011influence}. Timbre is particularly critical for differentiating emotions that may exhibit similar prosodic contours, such as anger and joy \cite{li2018contributions}. Emotionally salient speech often adopts stable, style specific articulatory postures, which in turn produce characteristic timbral shifts. For instance, tighter epilaryngeal narrowing in anger contrasts with more open, rounded settings in joy, producing distinct vocal profiles \cite{kim2020vocal}. Viewed over time and frequency, the log-magnitude spectrum forms an energy landscape whose local structure reflects the instantaneous resonances of the tract. These timbral variations can be quantified by features of the spectral envelope computed from the \textbf{log-magnitude spectrum} \cite{tzanetakis2002musical}.

    \textbf{Dynamic modulation} characterizes how emotion emerges from coordinated adjustments across the respiratory, phonatory, and articulatory systems \cite{scherer2003vocal}. These adjustments modulate overall energy flow and create temporal patterns of rhythm, stress, and pace. For example, higher arousal states such as excitement involve a stronger respiratory drive and faster articulatory movements, yielding sharp onsets and strong stress patterns. Conversely, lower arousal states exhibit smoother transitions and slower temporal organization \cite{alexander2025high}. We use the spectral flux (\textbf{log-magnitude rate}) to quantify this evolution. This measure captures the rate of change within the spectral energy landscape and is sensitive to vocal onsets, rapid energy shifts associated with stress, and transitions between phonetic segments. Consequently, log-magnitude rate provides a vocal correlate of the underlying physiological dynamics, effectively reflecting intensity and vitality \cite{bello2005tutorial}.

\subsubsection{Decomposing the Acoustic Signal}
    To link speech acoustics to physiological vocal state changes, we apply the short-time Fourier transform (STFT) to the waveform $x(t)$ and obtain a time-varying amplitude and phase field:
\begin{equation}
    S(t,\omega)=A(t,\omega)\,\mathrm{e}^{\mathrm{i}\,\phi(t,\omega)},
\end{equation}
    where $A(t,\omega)\!=\!|S(t,\omega)|$ is the magnitude spectrum, $\phi(t,\omega)\!=\!\arg S(t,\omega)$ is the phase spectrum, $t$ denotes time, and $\omega$ denotes angular frequency ($\mathrm{rad/s}$). Motivated by the preceding physiological analysis, we select the log-magnitude spectrum
\begin{equation}
    M(t,\omega)=\log A(t,\omega)=\log|S(t,\omega)|,
\end{equation}
    as the feature for encoding the vocal tract filter. It offers two advantages. First, log scaling compresses dynamic range and improves robustness to multiplicative channel effects (e.g., microphone gain and distance), yielding a stable spectral envelope that underlies timbre. Second, under this model $S(t,\omega)\approx \mathrm{G}(t,\omega)\mathrm{H}(t,\omega)$ (where $\mathrm{G}$ denotes the glottal source spectrum and $\mathrm{H}$ the vocal tract transfer function), magnitudes multiply and logs add, so $M=\log|S|\approx M_{\mathrm{G}}+M_{\mathrm{H}}$, with the slowly varying $M_{\mathrm{H}}$ corresponding to the vocal tract configuration, which makes the source and filter more separable.

    While $M(t,\omega)$ summarizes the instantaneous filter state, emotional expression is dynamic. To capture this temporal evolution, we define the log-magnitude rate:
\begin{equation}
    \rho(t,\omega)=\partial_t M(t,\omega),
\end{equation} 
    also known as spectral flux, which quantifies the rate of change within the spectral energy landscape. It is sensitive to acoustic onsets driven by increases in subglottal pressure, to rapid spectral changes caused by fast articulatory movements, and to the rhythm and pace of an utterance. Moreover, $\rho$ is invariant to constant recording-level changes: if a channel gain scales the magnitude by a factor $\kappa>0$ so that $A'(t,\omega)=\kappa A(t,\omega)$, then $M'(t,\omega)=M(t,\omega)+\log\kappa$ and hence $\partial_t M'=\partial_t M$, leaving $\rho$ unchanged. Thus, $M$ provides static timbre-related information, whereas $\rho$ provides complementary dynamic information that reflects the underlying physiological modulation.

    However, the features $M$ and $\rho$ are derived entirely from the magnitude of the signal, discarding the phase spectrum $\phi(t,\omega)$. The phase and its local gradients contain crucial, complementary information about the phonatory source and its interaction with the vocal tract filter. We therefore extract two additional features from the partial derivatives of the phase.

    To capture the dynamics of the phonatory source, we use the IF, defined as the time derivative of the phase:
\begin{equation}
    f_{\mathrm{inst}}(t,\omega)=\frac{1}{2\pi}\,\partial_t \phi(t,\omega),
\end{equation}
    $f_{\mathrm{inst}}$ measures the local oscillation rate within each time frequency region and serves as a direct acoustic correlate of laryngeal fold vibration. It captures the fundamental frequency contour underlying prosody and intonation as well as micro prosodic fluctuations linked to the phonation type. A global phase offset $\phi\mapsto \phi+C$ leaves $f_{\mathrm{inst}}$ unchanged.

    Group delay (GD) reveals nuances of articulation and interaction, measuring the negative frequency derivative of phase:
\begin{equation}
    \tau_{\mathrm{g}}(t,\omega) = -\,\partial_\omega \phi(t,\omega),
\end{equation}
    where $\tau_{\mathrm{g}}$ quantifies phase dispersion across frequency and is shaped by both vocal tract resonances and glottal pulse asymmetry. For minimum phase tract responses, $\tau_{\mathrm{g}}$ exhibits positive peaks near formant frequencies, complementing the magnitude-based envelope in $M$. When non-minimum phase effects are present, stabilized or modified forms of group delay preserve diagnostic value while mitigating spurious excursions. A global phase offset does not affect $\tau_{\mathrm{g}}$.

    Overall, we adopt the quartet $\{\,M,\ \rho=\partial_t M,\ f_{\mathrm{inst}}=\tfrac{1}{2\pi}\partial_t\phi,\ \tau_{\mathrm{g}}=-\partial_\omega\phi\,\}$ as a minimal physiology informed representation of vocal acoustics. It covers the full physiological pathway by aligning respiration with temporal energy flow, the laryngeal source with bandwise oscillation rate, and the vocal tract with the resonant envelope and phase dispersion. It further spans the physical structure of acoustic features, with $M$ anchoring the spectral envelope and timbre, $\rho$ capturing dynamic energy and onsets, $f_{\mathrm{inst}}$ encoding prosody and microprosody, and $\tau_{\mathrm{g}}$ revealing resonances, antiresonances, and glottal pulse asymmetry. Mathematically, it uses the STFT magnitude and phase together with their first derivatives in time and frequency, with explicit units and basic invariances: $\rho$ is unchanged under global gain, and $f_{\mathrm{inst}}$ and $\tau_{\mathrm{g}}$ are unchanged under global phase offsets. The representation is necessary, since removing any component leaves its associated physiological axis underdetermined and unrecoverable from the remaining terms without additional assumptions.

\subsection{Quaternion Spectrotemporal Encoder (QSE)}
    As discussed above, the quartet $\{M,\rho,f_{\mathrm{inst}},\tau_{\mathrm{g}}\}$ is a compact yet coupled representation: $M$ provides a stable timbral foundation, while $(\rho,f_{\mathrm{inst}},\tau_{\mathrm{g}})$ are derivative-based dynamics whose co-orientation at each $(t,\omega)$ carries discriminative features. Hence the extractor must encode their joint structure. We require an operator that (i) mixes the scalar and vector parts bilinearly with the antisymmetric sign pattern dictated by AM–FM and acoustic physics, (ii) is equivariant to rotations within the dynamics subspace $(\rho,f_{\mathrm{inst}},\tau_{\mathrm{g}})\mapsto R(\rho,f_{\mathrm{inst}},\tau_{\mathrm{g}})$, so that learned patterns are unaffected by arbitrary reparameterizations of these three physically homogeneous axes under the same timbral baseline, and (iii) ties cross-channel parameters to preserve these symmetries and reduce sample complexity. Hamilton-structured quaternion convolutions satisfy these properties. We therefore introduce the Quaternion Spectrotemporal Encoder (QSE), which embeds the four signals into a spectrotemporal field $\mathcal{Q}(t,\omega)$ as:
\begin{equation}
    \mathcal{Q}(t,\omega) 
    = \underbrace{M(t,\omega)}_{r} 
    + \mbi\underbrace{\rho(t,\omega)}_{a} 
    + \mbj\underbrace{f_{\mathrm{inst}}(t,\omega)}_{b}  
    + \mbk\underbrace{\tau_{\mathrm{g}}(t,\omega)}_{c},
    \label{eq:q-field}
\end{equation}
    This assignment respects units and invariances while concentrating timbre in the scalar part and dynamics in the vector part. The QSE processes the input field $\mathcal{Q}^{(0)}$ using a stack of $L$ identical blocks that preserve the Hamilton coupling throughout. Each block maps an input $\mathcal{Q}^{(l)}$ to an output $\mathcal{Q}^{(l+1)}$ via a sequence of quaternion operations:
\begin{equation}
    \label{eq:q-block}
    \mathcal{Q}^{(l+1)}
    \!=\!
    \operatorname{Pool}
    \!\left(
    \mathrm{q\text{-}ReLU}
    \big(
    \mathrm{QBN}\big(\mathcal{H}(\mathcal{W}^{(l)})\ast \mathcal{Q}^{(l)} + \mathcal{B}^{(l)}\big)
    \big)
    \right),
\end{equation}
    where $\ast$ denotes convolution, $\mathcal{B}^{(l)}$ is a quaternion bias, and $\mathcal{H}(\mathcal{W}^{(l)})$ is the Hamilton product. In the first block, the components of $\mathcal{Q}^{(0)}$ correspond directly to the input features, allowing the filters to capture local AM-FM \cite{DBLP:journals/tsp/MaragosKQ93a} and signals co-variations. In deeper blocks, they operate on latent quaternion features governed by the same Hamilton product. The key components of this block are detailed below.

    The core operation is the quaternion convolution. Let
\begin{align}
\mbx = [\, M,\ \rho,\ f_{\mathrm{inst}},\ \tau_{\mathrm{g}} \,]^{\top}, 
\
\mby = [\, Y_{\mathrm{M}},\ Y_{\rho},\ Y_{f},\ Y_{\tau} \,]^{\top}.
\end{align}
    and parameterize the kernels by 4 shared real filters $\{\mathcal{W}_{\mathrm{M}},\mathcal{W}_{\rho},\mathcal{W}_{f},\mathcal{W}_{\tau}\}$. The induced Hamilton product is
\begin{equation}
\label{hamilton}
    \mathcal{H}(\mathcal{W})=
    \begin{bmatrix}
    \mathcal{W}_{\mathrm{M}} & -\mathcal{W}_{\rho} & -\mathcal{W}_{f} & -\mathcal{W}_{\tau}\\
    \mathcal{W}_{\rho} & \mathcal{W}_{\mathrm{M}} & -\mathcal{W}_{\tau} &\mathcal{W}_{f}\\
    \mathcal{W}_{f} & \mathcal{W}_{\tau} & \mathcal{W}_{\mathrm{M}} & -\mathcal{W}_{\rho}\\
    \mathcal{W}_{\tau} & -\mathcal{W}_{f} & \mathcal{W}_{\rho} & \mathcal{W}_{\mathrm{M}}
    \end{bmatrix},
\end{equation}
    and the quaternion convolution is:
    \(\mby=\mathcal{H}(\mathcal{W})\ast\mbx.\)

    This construction implements the Hamilton product, ensuring that each output is a structured mixture of all inputs. This induces a physical relational inductive bias that encodes source-filter covariation and energy phase interplay by design and yields parameter efficiency.

    To stabilize deep training, we employ specialized normalization and activation functions. The Quaternion Batch Normalization (QBN) operates on each of the four components independently with learnable affine parameters:
{\small
\begin{equation}
\label{eq:qbn}
    \begin{split}
    \mathrm{QBN}(\mathcal{Q}) ={}& (\gamma_r\,\mathrm{BN}(r)+\beta_r)
    + \mbi\,(\gamma_a\,\mathrm{BN}(a)+\beta_a) \\
    &+ \mbj\,(\gamma_b\,\mathrm{BN}(b)+\beta_b) 
    + \mbk\,(\gamma_c\,\mathrm{BN}(c)+\beta_c)\,.
    \end{split}
\end{equation}
}

    Since $r=M$ and $(a,b,c)=(\rho,f_{\mathrm{inst}},\tau_{\mathrm{g}})$ have different units and dynamic ranges, this choice avoids unit mismatch, and the Hamilton mixing in the next convolution restores cross component coupling. The subsequent radial nonlinearity is
\begin{equation}
\label{eq:qrelu}
    \mathrm{qReLU}(\mbq)
    = \frac{\max(\|\mbq\|,0)}{\|\mbq\|+\varepsilon}\,\mbq,
    \|\mbq\|=\sqrt{r^2+a^2+b^2+c^2},
\end{equation}
    which rescales only the magnitude while leaving the quaternion direction unchanged, thereby preserving the relative orientation among $M$, $\rho$, $f_{\mathrm{inst}}$, and $\tau_{\mathrm{g}}$ that encodes phase related structure and suppresses degenerate low energy regions.

    Finally, we apply $\operatorname{MaxPool}$ with a short window and stride along the frequency axis only. This enhances spectral features correlated with emotion while preserving full temporal resolution for rhythm and prosody.

    After the final QSE block, the quaternion feature map $\mathcal{Q}^{(L)}$ is represented as four real channels. To create a sequence compatible with downstream models, we permute the axes and vectorize the spectral channel coefficients at each time step, producing a time-indexed embedding sequence:
\begin{equation}
\label{eq:q-reshape}
    \mbG
    =
    \begin{bmatrix}
    \operatorname{vec}\!\big(\mathcal{Q}^{(L)}(\cdot,\cdot,1)\big)^\top\\
    \vdots\\
    \operatorname{vec}\!\big(\mathcal{Q}^{(L)}(\cdot,\cdot,T)\big)^\top
    \end{bmatrix}
    \in \bbR^{T\times D},
\end{equation}    
    where $T$ is the padded sequence length in \Cref{problem}, and $\operatorname{vec}(\cdot)$ stacks, at each time $t$, all spectral channel coefficients across the four real components into a single vector of length $D$. This final representation encapsulates the learned spectrotemporal structure in a fixed dimensional vector.

\subsection{Utterance Level Feature Aggregation}

    In the latent speech representation branch (\Cref{structure}(a)), a frozen, general-purpose SSL backbone extracts a frame wise sequence of latent representations from the raw waveform, denoted as $\mbE=\{\mathbf{e}_t\}_{t=1}^{T_e}$. We then apply a lightweight latent transformation module to adapt latent features:
\begin{equation}
\mathbf{e}'_t=\sigma!\big(\mathcal{W}_e,\mathbf{e}_t+\mathbf{b}_e\big), \ \sigma=\mathrm{GELU}.
\end{equation}

    In the vocal feature representation branch (\Cref{structure}(b)), following the VAP-informed feature decomposition and the QSE encoding described above, we obtain a time-indexed embedding sequence $\mbG=\{\mathbf{g}_t\}_{t=1}^{T_g}$ that captures physiology informed emotional cues. Here, $T_e$ and $T_g$ denote the sequence lengths of the latent and vocal streams, respectively. To preserve the QSE structured amplitude-phase coupling, we feed the QSE stream directly to the utterance level aggregator.

    To obtain utterance level features while retaining transient emotional patterns, we summarize each sequence using masked attention pooling with the binary mask defined in \Cref{problem}. We illustrate the computation on $\mbG=\{\mathbf{g}_t\}_{t=1}^{T_g}$, and the same procedure applies to the latent branch. We first compute frame wise scores \(s_t=\mathbf{w}^{\!\top}\mathbf{g}_t+b,\) and apply masking:
\begin{equation}
    \hat s_t =
    \begin{cases}
    s_t, & (\mbm)_t=1,\\
    -\infty, & (\mbm)_t=0.
    \end{cases}
    \alpha_t=\frac{\exp(\hat s_t)}{\sum_{j=1}^{T_g}\exp(\hat s_j)}, \ \sum_{t=1}^{T_g}\alpha_t=1,
\end{equation}
    $s_t$ is the frame score, and $\alpha_t$ is the attention weight. Masked positions are implemented as a large negative constant so that their weights are zero. The attention-pooled sentence vector is
\begin{equation}
   \mathbf{z}=\sum_{t=1}^{T_g}\alpha_t\,\mathbf{g}_t.
\end{equation}

    Applying the same operator to both branches gives
\begin{equation}
    \mathbf{z}_{\mathrm{latent}}=\mathcal{A}\big(\{\mathbf{e}'_t\},\,\mbm^{(e)}\big),\qquad
    \mathbf{z}_{\mathrm{vocal}}=\mathcal{A}\big(\{\mathbf{g}_t\},\,\mbm^{(g)}\big),
\end{equation}
    where $\mathcal{A}$ denotes the masked attention pooling defined above, and $\mbm^{(e)}$, $\mbm^{(g)}$ are the branch masks derived from $T_e$ and $T_g$.

    Both vectors $\mathbf{z}_{\mathrm{latent}}$ and $\mathbf{z}_{\mathrm{vocal}}$ encode core constituents of speech but differ in parameterization and constraints. The vocal representation branch explicitly factorizes $(M,\rho,f_{\mathrm{inst}},\tau_{\mathrm{g}})$ and preserves phase gradient structure through the Hamilton product, thus retaining features that generic magnitude based encoders rarely preserve. The backbone path is learned end-to-end on large corpora with wide receptive fields, yielding rich texture and context while mainly capturing amplitude driven structure such as spectral envelopes, onsets, and spectral shapes, whereas absolute phase is often down-weighted by Mel smoothing and temporal striding.

    Physiologically, changes in arousal and valence alter subglottal pressure, laryngeal tension, and closure quotient, thereby shifting $f_{\mathrm{inst}}$ trajectories and tightening or dispersing phase relations as measured by $\tau_{\mathrm{g}}$. While the magnitude may remain stable as interharmonic phase varies, phase gradients provide emotion-specific details that complement magnitude-driven features. Mathematically, the QSE formulation corresponds to AM–FM decomposition with physically motivated coupling, which improves identifiability under gain and phase shifts. In parallel, the latent representation branch contributes a general representation and long range dependencies not targeted by QSE.

    The two representations are complementary yet partially overlapping. Overlap in magnitude-driven structure provides robustness through agreement, whereas phase-based residuals in $\mathbf{z}_{\mathrm{vocal}}$ add discriminative power. The masked attention pooling defined above, together with the subsequent contrastive projection, aligns shared components into a common subspace while preserving branch-specific residuals. Consequently, combining $\mathbf{z}_{\mathrm{latent}}$ and $\mathbf{z}_{\mathrm{vocal}}$ improves identifiability while preserving physical interpretability and robustness.

\subsection{Contrastive Projection and Alignment (CPA) Framework}
    These two branches provide complementary utterance level representations, $\mathbf{z}_{\mathrm{vocal}}$ carries a VAP-informed emotional vocal analysis, and $\mathbf{z}_{\mathrm{latent}}$ carries general latent speech regularities using an SSL model. Although both originated from the same human voice, their distinct expressions mean they reside in disparate feature spaces, which hinders effective fusion. We therefore introduce the CPA framework to align these two representations in a shared feature space.

    The alignment through a lightweight, symmetric contrastive objective. Both utterance level vectors are mapped into a common $d_{\mathrm{align}}$ dimensional space and $\ell_2$ normalized:
\begin{equation}
    \mathbf{u}_b=\frac{\mathcal{W}_2^{(\mathrm{q})}\,\sigma\!\big(\mathcal{W}_1^{(\mathrm{q})}\,\mathbf{z}_{\mathrm{vocal}}^{(b)}\big)}{\big\|\mathcal{W}_2^{(\mathrm{q})}\,\sigma\!\big(\mathcal{W}_1^{(\mathrm{q})}\,\mathbf{z}_{\mathrm{vocal}}^{(b)}\big)\big\|_2},
    \mathbf{v}_b=\frac{\mathcal{W}_2^{(\mathrm{s})}\,\sigma\!\big(\mathcal{W}_1^{(\mathrm{s})}\,\mathbf{z}_{\mathrm{latent}}^{(b)}\big)}{\big\|\mathcal{W}_2^{(\mathrm{s})}\,\sigma\!\big(\mathcal{W}_1^{(\mathrm{s})}\,\mathbf{z}_{\mathrm{latent}}^{(b)}\big)\big\|_2},
\end{equation}
   where $b\in\{1,\dots,B\}$ indexes utterances in a minibatch;  $(\mathrm{q})$ and $(\mathrm{s})$ denote branch-specific parameters for the vocal and latent branches, $\mathcal{W}_1^{(\cdot)},\mathcal{W}_2^{(\cdot)}$ are small trainable maps, $\sigma$ is a pointwise nonlinearity, and $\|\cdot\|_2$ denotes the Euclidean norm.
    
    Let $s_{b,b'}=\mathbf{u}_b^{\top}\mathbf{v}_{b'}$ and let $\eta>0$ be the temperature. The bidirectional losses are:
\begin{align}
\label{eq:cpaf:infonce}
    \mathcal{L}_{\mathrm{Q}\rightarrow \mathrm{S}}
    &=\frac{1}{B}\sum_{b=1}^{B}
    -\log\frac{\exp(s_{b,b}/\eta)}{\sum_{b'=1}^{B}\exp(s_{b,b'}/\eta)},\\
    \mathcal{L}_{\mathrm{S}\rightarrow \mathrm{Q}}
    &=\frac{1}{B}\sum_{b=1}^{B}
    -\log\frac{\exp(s_{b,b}/\eta)}{\sum_{b'=1}^{B}\exp(s_{b',b}/\eta)}.
\label{infonce2}
\end{align}

    We treat $(\mathbf{u}_b,\mathbf{v}_b)$ from the same utterance as the positive pair, and all $(\mathbf{u}_b,\mathbf{v}_{b'})$ with $b'\neq b$ as negatives. The objective is to pull positives together in the shared space and push negatives apart. This choice aligns views at the same utterance granularity while avoiding class level shortcuts: positives share utterance prosodic but differ by branch specific inductive biases, so pulling them together transfers invariances without erasing complementary features; negatives drawn from other utterances enforce inter-utterance margins and prevent collapse. The symmetric form enforces mutual predictability, and $\eta$ provides an adaptive angular margin.

    During CPA framework pre-training, the SSL model is frozen. The vocal feature representation branch and both projection heads are trainable, and the objective is $\mathcal{L}_{\mathrm{CPA}}=\tfrac{1}{2}\big(\mathcal{L}_{\mathrm{Q}\rightarrow \mathrm{S}}+\mathcal{L}_{\mathrm{S}\rightarrow \mathrm{Q}}\big)$. This transfers the latent speech representation invariances of the SSL model to the VAP-informed pathway while preserving the latter’s inductive bias.

\subsection{Fusion and Classification Head}
    Following the CPA pre-training, the two branches yield aligned embeddings $\mathbf{z}_{\mathrm{latent}}$ and $\mathbf{z}_{\mathrm{vocal}}$ from utterance level aggregation. We design a fusion head to model the final interactions that produce the emotion classification, without reintroducing heavy parameterization or breaking the invariances established upstream. We therefore adopt a shallow Transformer encoder followed by a compact classifier.

    We first stack the two representations as
   \( \mathcal{Z}^{(0)}=\big[\ \mathbf{z}_{\mathrm{latent}}\ \|\ \mathbf{z}_{\mathrm{vocal}}\ \big]\),
    where the row order is fixed and $[\cdot\ \|\ \cdot]$ denotes stacking along the sequence axis. We then use a shallow Transformer encoder to fuse the sequence:
\begin{equation}
\label{eq:fusion-encoder}
    \begin{aligned}
    \mathcal{U}^{(\ell)} &= \mathrm{MHA}\!\big(\mathrm{LN}(\mathcal{Z}^{(\ell)})\big) + \mathcal{Z}^{(\ell)},\\
    \mathcal{Z}^{(\ell+1)} &= \mathrm{FFN}\!\big(\mathrm{LN}(\mathcal{U}^{(\ell)})\big) + \mathcal{U}^{(\ell)}, \ell=0,1,\dots,L-1,
    \end{aligned}
\end{equation}
    where $\mathrm{LN}$ is layer normalization, $\mathrm{MHA}$ is multi-head self-attention with scaled dot product, and $\mathrm{FFN}$ is two-layer positionwise network with a pointwise nonlinearity (GELU). This process implements an expressive yet parameter efficient bilinear mixing between modalities that subsumes concatenation and learned gating as special cases. It is also scale insensitive, and phase offset invariances are retained, since attention acts on normalized embeddings and depends on angular relations.

    The encoder output $\mathcal{Z}^{(L)}=[\,\mathbf{h}^{(L)}_{\mathrm{latent}},\ \mathbf{h}^{(L)}_{\mathrm{vocal}}\,]$ contains tokens. We take the first token as the fused representation:
    \(\mathbf{z}_{\mathrm{fuse}}=\mathbf{h}^{(L)}_{\mathrm{latent}}\),
    which functions as a fixed query that aggregates information from both inputs via \Cref{eq:fusion-encoder}. A compact post-network produces class logits:
\begin{equation}
\label{eq:cls-head}
    \mathbold{\ell} = \mathcal{W}_{2}\,\sigma\!\big(\mathcal{W}_{1}\,\mathbf{z}_{\mathrm{fuse}}+\mathbf{b}_{1}\big)+\mathbf{b}_{2}\in\bbR^{C},
\end{equation}
    with $\sigma$ a ReLU and $C$ the number of emotion classes. In this stage, the supervised objective is the cross-entropy Loss:
\begin{equation}
\label{eq:ce}
    \mathcal{L}_{\mathrm{CE}}=-\log\frac{\exp(\ell_{y})}{\sum_{c=1}^{C}\exp(\ell_{c})}.
\end{equation}

\section{Experiments}
\label{sec:experiments}
    In this section, we present a comprehensive empirical evaluation of PhysioSER. We begin by describing the experimental setup, including datasets, the evaluation protocol, and implementation details. We then evaluate the effectiveness of PhysioSER when paired with various SSL backbones and different languages. Finally, we conduct ablation studies to quantify the contribution of each key component. 


\sisetup{round-mode=places,round-precision=2,table-number-alignment=center,detect-weight=true,detect-inline-weight=math}
\sisetup{detect-weight,retain-explicit-plus,table-number-alignment=center}

\begin{table*}[t]
    \centering
    \caption{Results on English datasets. $\Delta(\%)$ is the relative improvement of Ours over the backbone-only on the same dataset.}
    \label{tab:multi_ds_ssl}
    \setlength{\tabcolsep}{4.5pt}
    \renewcommand{\arraystretch}{0.8}
    \scriptsize
    \begin{threeparttable}
    \begin{adjustbox}{width=0.95\textwidth}
    \begin{tabular}{
    l  l |
    S[table-format=+3.2, table-column-width=1.3cm]
    S[table-format=+3.2, table-column-width=1.3cm]
    S[table-format=+3.2, table-column-width=0.9cm] |
    S[table-format=+3.2, table-column-width=1.3cm]
    S[table-format=+3.2, table-column-width=1.3cm]
    S[table-format=+3.2, table-column-width=0.9cm] |
    S[table-format=+3.2, table-column-width=1.3cm]
    S[table-format=+3.2, table-column-width=1.3cm]
    S[table-format=+3.2, table-column-width=0.9cm] 
    }
    \toprule
    \multirow{2}{*}{Dataset}  &  \multirow{2}{*}{Backbone}
    &  \multicolumn{3}{c |}{\wa~(\%)}  
    &  \multicolumn{3}{c |}{\ua~(\%)}  
    &  \multicolumn{3}{c}{\mf~(\%)}  \\
    \cmidrule(lr){3-5}\cmidrule(lr){6-8}\cmidrule(lr){9-11}
    &  &  {Backbone-only}  &  {Ours}  &  {$\Delta$}  
      &  {Backbone-only}  &  {Ours}  &  {$\Delta$}
      &  {Backbone-only}  &  {Ours}  &  {$\Delta$}  \\
    \midrule
    \multirow{6}{*}{\shortstack{ {CREMA-D} \\ (6 classes)}}
    & WavLM
      & 69.69 & \textbf{75.20} & +7.91
      & 70.22 & \textbf{75.62} & +7.69
      & 69.79 & \textbf{75.63} & +8.37 \\
    & HuBERT
      & 66.13 & \textbf{69.89} & +5.69
      & 66.68 & \textbf{70.33} & +5.47
      & 66.33 & \textbf{69.51} & +4.79 \\
    & Wav2vec2
      & 57.58 & \textbf{64.38} & +11.81
      & 58.13 & \textbf{64.84} & +11.54
      & 56.98 & \textbf{64.13} & +12.55 \\
    & Emotion2Vec
      & 69.02 & \textbf{72.98} & +5.74
      & 69.52 & \textbf{73.34} & +5.49
      & 69.16 & \textbf{72.99} & +5.54 \\
    & BEATs
      & 71.51 & \textbf{75.08} & +4.99
      & 71.78 & \textbf{75.59} & +5.31
      & 71.67 & \textbf{75.07} & +4.74 \\
    & CLAP
      & 53.90 & \textbf{59.81} & +10.96
      & 54.19 & \textbf{59.84} & +10.43
      & 53.72 & \textbf{59.35} & +10.48 \\
    \midrule
    \multirow{6}{*}{\shortstack{ {EMNS} \\ (8 classes)}}
    &  WavLM
      & 77.97 & \textbf{83.05} & +6.52
      & 77.96 & \textbf{82.73} & +6.12
      & 77.47 & \textbf{82.44} & +6.42 \\
    &  HuBERT
      & 40.25 & \textbf{62.29} & +54.76
      & 40.76 & \textbf{63.24} & +55.15
      & 37.61 & \textbf{63.25} & +68.17 \\
    &  Wav2vec2
      & 32.63 & \textbf{59.32} & +81.80
      & 33.35 & \textbf{60.60} & +81.71
      & 32.61 & \textbf{58.64} & +79.82 \\
    &  Emotion2Vec
      & 63.98 & \textbf{67.80} & +5.97
      & 63.97 & \textbf{67.40} & +5.36
      & 63.74 & \textbf{66.91} & +4.97 \\
    &  BEATs
      & 81.36 & \textbf{85.59} & +5.20
      & 81.25 & \textbf{85.63} & +5.39
      & 81.02 & \textbf{85.17} & +5.12 \\
    &  CLAP
      & 64.15 & \textbf{80.51} & +25.50
      & 65.27 & \textbf{81.32} & +24.59
      & 65.68 & \textbf{80.10} & +21.95 \\
    \midrule
    \multirow{6}{*}{\shortstack{ JLCorpus \\ (10 classes)}}
    &  WavLM
      & 73.54 & \textbf{84.17} & +14.45
      & 72.95 & \textbf{83.75} & +14.80
      & 72.20 & \textbf{83.64} & +15.84 \\
    &  HuBERT
      & 51.25 & \textbf{71.88} & +40.25
      & 50.32 & \textbf{71.20} & +41.49
      & 49.28 & \textbf{70.96} & +43.99 \\
    &  Wav2vec2
      & 28.33 & \textbf{66.88} & +136.07
      & 28.14 & \textbf{66.74} & +137.17
      & 24.86 & \textbf{66.31} & +166.73 \\
    &  Emotion2Vec
      & 68.75 & \textbf{80.63} & +17.28
      & 68.50 & \textbf{80.25} & +17.15
      & 67.96 & \textbf{79.91} & +17.58 \\
    &  BEATs
      & 73.13 & \textbf{86.46} & +18.23
      & 72.22 & \textbf{86.86} & +20.27
      & 71.87 & \textbf{86.41} & +20.22 \\
    &  CLAP
      & 50.00 & \textbf{76.25} & +52.50
      & 49.96 & \textbf{76.18} & +52.48
      & 48.28 & \textbf{76.31} & +58.06 \\
    \midrule
    \multirow{6}{*}{\shortstack{ {EmoV-DB} \\ (5 classes)}}
    &  WavLM
      & 98.11 & \textbf{99.42} & +1.34
      & 98.03 & \textbf{99.34} & +1.34
      & 97.96 & \textbf{99.41} & +1.48 \\
    &  HuBERT
      & 95.79 & \textbf{98.19} & +2.51
      & 95.43 & \textbf{98.16} & +2.86
      & 95.44 & \textbf{98.07} & +2.76 \\
    &  Wav2vec2
      & 81.44 & \textbf{97.10} & +19.23
      & 80.56 & \textbf{96.95} & +20.35
      & 80.52 & \textbf{96.92} & +20.37 \\
    &  Emotion2Vec
      & 95.57 & \textbf{99.27} & +3.87
      & 95.33 & \textbf{99.19} & +4.05
      & 95.30 & \textbf{99.22} & +4.11 \\
    &  BEATs
      & 98.77 & \textbf{99.20} & +0.44
      & 98.64 & \textbf{99.15} & +0.52
      & 98.66 & \textbf{99.17} & +0.52 \\
    &  CLAP
      & 95.94 & \textbf{97.90} & +2.04
      & 95.80 & \textbf{97.55} & +1.83
      & 95.64 & \textbf{97.73} & +2.19 \\
    \midrule
    \multirow{6}{*}{\shortstack{ {RAVDESS} \\ (8 classes)}}
    &  WavLM
      & 66.31 & \textbf{84.72} & +27.76
      & 64.19 & \textbf{84.71} & +31.97
      & 64.02 & \textbf{84.02} & +31.24 \\
    &  HuBERT
      & 59.03 & \textbf{65.28} & +10.59
      & 56.45 & \textbf{65.12} & +15.36
      & 56.63 & \textbf{64.34} & +13.61 \\
    &  Wav2vec2
      & 44.44 & \textbf{48.61} & +9.38
      & 42.48 & \textbf{47.46} & +11.72
      & 40.16 & \textbf{47.03} & +17.11 \\
    &  Emotion2Vec
      & 62.15 & \textbf{81.60} & +31.30
      & 58.10 & \textbf{81.10} & +39.59
      & 57.57 & \textbf{81.45} & +41.48 \\
    &  BEATs
      & 81.60 & \textbf{82.99} & +1.70
      & 78.89 & \textbf{83.05} & +5.27
      & 79.07 & \textbf{82.03} & +3.74 \\
    &  CLAP
      & 57.64 & \textbf{68.75} & +19.27
      & 53.63 & \textbf{67.42} & +25.71
      & 53.83 & \textbf{67.21} & +24.86 \\
    \midrule
    \bottomrule
    \end{tabular}
    \end{adjustbox}

    \end{threeparttable}
\end{table*}

\begin{table*}[!ht]
    \centering
    \scriptsize
    \caption{Results on non-English datasets. $\Delta$ is the relative improvement of Ours over the Backbone-only on the same dataset.}
    \label{tab:multi_ds_ssl_en}
    \setlength{\tabcolsep}{4.5pt}
    \renewcommand{\arraystretch}{0.8}
    \begin{threeparttable}
    \begin{adjustbox}{width=0.98\textwidth}
    \begin{tabular}{l l |
    S[table-format=+3.2, table-column-width=1.3cm]
    S[table-format=+3.2, table-column-width=1.3cm]
    S[table-format=+3.2, table-column-width=0.9cm] |
    S[table-format=+3.2, table-column-width=1.3cm]
    S[table-format=+3.2, table-column-width=1.3cm]
    S[table-format=+3.2, table-column-width=0.9cm] |
    S[table-format=+3.2, table-column-width=1.3cm]
    S[table-format=+3.2, table-column-width=1.3cm]
    S[table-format=+3.2, table-column-width=0.9cm] 
    }
    \toprule
    \multirow{2}{*}{Dataset} & \multirow{2}{*}{Backbone}
    & \multicolumn{3}{c|}{\wa~(\%)}  
    & \multicolumn{3}{c|}{\ua~(\%)}  
    & \multicolumn{3}{c}{\mf~(\%)} \\
    \cmidrule(lr){3-5}\cmidrule(lr){6-8}\cmidrule(lr){9-11}
    & & {Backbone-only} & {Ours} & {$\Delta$}
      & {Backbone-only} & {Ours} & {$\Delta$}
      & {Backbone-only} & {Ours} & {$\Delta$} \\
    \midrule
    \multirow{6}{*}{\shortstack{ {RESD} \\ (Russian) \\(7 classes)}}
    &  WavLM     & 50.00 & \textbf{59.64} & +19.28  & 48.98 & \textbf{58.97} & +20.40  & 48.78 & \textbf{58.01} & +18.92 \\
    &  HuBERT    & 34.29 & \textbf{43.95} & +28.17  & 32.33 & \textbf{43.30} & +33.93  & 29.03 & \textbf{40.91} & +40.92 \\
    &  Wav2vec2   & 25.00 & \textbf{41.70} & +66.80  & 22.92 & \textbf{40.75} & +77.79  & 16.79 & \textbf{39.45} & +134.96 \\
    &  Emotion2Vec   & 43.93 & \textbf{51.57} & +17.39  & 42.86 & \textbf{50.59} & +18.04  & 42.66 & \textbf{50.12} & +17.49 \\
    &  BEATs     & 61.43 & \textbf{71.75} & +16.80  & 60.79 & \textbf{70.83} & +16.52  & 60.86 & \textbf{70.65} & +16.09 \\
    &  CLAP      & 45.36 & \textbf{54.71} & +20.61  & 43.80 & \textbf{54.18} & +23.70  & 42.66 & \textbf{53.69} & +25.86 \\
    \midrule
    \multirow{6}{*}{\shortstack{ {nEMO} \\ (Polish) \\(6 classes)}}
    &  WavLM     & 75.00 & \textbf{90.63} & +20.84  & 74.57 & \textbf{90.56} & +21.44  & 74.44 & \textbf{90.57} & +21.67 \\
    &  HuBERT    & 66.85 & \textbf{82.92} & +24.04  & 66.07 & \textbf{82.74} & +25.23  & 65.74 & \textbf{82.63} & +25.69 \\
    &  Wav2vec2   & 37.61 & \textbf{77.46} & +105.96 & 36.70 & \textbf{76.76} & +109.16 & 33.93 & \textbf{77.24} & +127.65 \\
    &  Emotion2Vec   & 77.01 & \textbf{86.83} & +12.75  & 76.61 & \textbf{86.40} & +12.78  & 76.70 & \textbf{86.68} & +13.01 \\
    &  BEATs     & 86.16 & \textbf{93.64} & +8.68   & 85.74 & \textbf{93.40} & +8.93   & 85.77 & \textbf{93.48} & +8.99  \\
    &  CLAP      & 66.63 & \textbf{85.60} & +28.47  & 66.20 & \textbf{85.36} & +28.94  & 66.02 & \textbf{85.48} & +29.48 \\
    \midrule
    \multirow{6}{*}{\shortstack{ {AESDD} \\ (Greek) \\(5 classes)}}
    &  WavLM     & 75.00 & \textbf{89.17} & +18.89  & 76.76 & \textbf{88.93} & +15.85  & 86.83 & \textbf{98.10} & +12.98 \\
    &  HuBERT    & 69.17 & \textbf{73.33} & +6.01   & 70.48 & \textbf{73.53} & +4.33   & 68.35 & \textbf{73.34} & +7.30  \\
    &  Wav2vec2  & 34.17 & \textbf{67.50} & +97.54  & 35.27 & \textbf{66.89} & +89.65  & 31.06 & \textbf{66.71} & +114.78 \\
    &  Emotion2Vec   & 64.17 & \textbf{91.67} & +42.85  & 65.38 & \textbf{91.80} & +40.41  & 64.03 & \textbf{91.67} & +43.17 \\
    &  BEATs     & 83.33 & \textbf{88.33} & +6.00   & 83.80 & \textbf{87.85} & +4.83   & 83.78 & \textbf{87.66} & +4.63  \\
    &  CLAP      & 58.33 & \textbf{72.50} & +24.29  & 59.33 & \textbf{72.26} & +21.79  & 57.89 & \textbf{72.69} & +25.57 \\
    \midrule
    \multirow{6}{*} {\shortstack{ {Emozionalmente} \\ (Italian) \\(7 classes)}}
    &  WavLM     & 68.04 & \textbf{72.17} & +6.07   & 68.18 & \textbf{71.98} & +5.57   & 67.98 & \textbf{72.01} & +5.93  \\
    &  HuBERT    & 56.58 & \textbf{58.30} & +3.04   & 56.73 & \textbf{59.19} & +4.34   & 55.40 & \textbf{57.97} & +4.64  \\
    &  Wav2vec2   & 36.16 & \textbf{45.58} & +26.05  & 36.57 & \textbf{45.49} & +24.39  & 34.83 & \textbf{45.62} & +30.98 \\
    &  Emotion2Vec   & 59.93 & \textbf{68.33} & +14.02  & 60.15 & \textbf{67.98} & +13.02  & 60.08 & \textbf{67.90} & +13.02 \\
    &  BEATs     & 67.25 & \textbf{68.19} & +1.40   & 67.38 & \textbf{67.97} & +0.88   & 67.21 & \textbf{67.99} & +1.16  \\
    &  CLAP      & 46.16 & \textbf{49.78} & +7.84   & 46.01 & \textbf{49.71} & +8.04   & 45.74 & \textbf{49.75} & +8.77  \\
    \midrule
    \multirow{6}{*}{\shortstack{ {Oréau} \\ (French) \\(7 classes)}}
    &  WavLM     & 58.00 & \textbf{84.00} & +44.83  & 60.67 & \textbf{84.05} & +38.54  & 58.83 & \textbf{83.78} & +42.41 \\
    &  HuBERT    & 46.00 & \textbf{58.00} & +26.09  & 49.09 & \textbf{60.55} & +23.34  & 45.22 & \textbf{58.40} & +29.15 \\
    &  Wav2vec2  & 28.00 & \textbf{44.00} & +57.14  & 30.50 & \textbf{45.35} & +48.69  & 23.38 & \textbf{44.62} & +90.85 \\
    &  Emotion2Vec   & 45.00 & \textbf{77.00} & +71.11  & 46.24 & \textbf{78.38} & +69.51  & 43.77 & \textbf{78.35} & +79.00 \\
    &  BEATs     & 51.00 & \textbf{52.00} & +1.96   & 54.02 & \textbf{55.46} & +2.67   & 50.79 & \textbf{51.50} & +1.40  \\
    &  CLAP      & 36.00 & \textbf{46.00} & +27.78  & 36.37 & \textbf{46.60} & +28.12  & 31.75 & \textbf{45.46} & +43.18 \\
    \midrule
    \multirow{6}{*}{\shortstack{{MESD} \\ (Spanish) \\(6 classes)}}
    &  WavLM     & 52.91 & \textbf{72.09} & +36.25  & 51.20 & \textbf{71.73} & +40.10  & 51.16 & \textbf{71.36} & +39.48 \\
    &  HuBERT    & 46.51 & \textbf{63.37} & +36.25  & 44.30 & \textbf{63.82} & +44.06  & 43.94 & \textbf{63.51} & +44.54 \\
    &  Wav2vec2   & 34.88 & \textbf{62.21} & +78.35  & 33.66 & \textbf{61.97} & +84.11  & 33.34 & \textbf{61.83} & +85.45 \\
    &  Emotion2Vec   & 63.95 & \textbf{70.35} & +10.01  & 63.19 & \textbf{70.48} & +11.54  & 62.81 & \textbf{69.42} & +10.52 \\
    &  BEATs     & 74.42 & \textbf{87.79} & +17.97  & 73.97 & \textbf{88.01} & +18.98  & 73.66 & \textbf{87.74} & +19.11 \\
    &  CLAP      & 65.70 & \textbf{67.44} & +2.65   & 64.15 & \textbf{68.41} & +6.64   & 63.78 & \textbf{67.21} & +5.38  \\
    \midrule
    \multirow{6}{*}{\shortstack{{SUBESCO} \\ (Bengali) \\(7 classes)}}
    &  WavLM     & 78.07 & \textbf{91.93} & +17.75  & 78.24 & \textbf{91.66} & +17.16  & 78.29 & \textbf{91.74} & +17.18 \\
    &  HuBERT    & 67.64 & \textbf{82.29} & +21.66  & 67.94 & \textbf{82.10} & +20.84  & 67.65 & \textbf{82.11} & +21.37 \\
    &  Wav2vec2   & 43.21 & \textbf{60.86} & +40.85  & 43.53 & \textbf{60.79} & +39.65  & 41.95 & \textbf{61.14} & +45.74 \\
    &  Emotion2Vec   & 75.07 & \textbf{87.79} & +16.94  & 75.07 & \textbf{87.58} & +16.66  & 74.88 & \textbf{87.61} & +17.00 \\
    &  BEATs     & 86.86 & \textbf{93.00} & +7.07   & 86.93 & \textbf{92.86} & +6.82   & 86.79 & \textbf{92.83} & +6.96  \\
    &  CLAP      & 63.93 & \textbf{82.93} & +29.72  & 64.23 & \textbf{82.75} & +28.83  & 62.49 & \textbf{82.69} & +32.33 \\
    \midrule
    \multirow{6}{*}{\shortstack{{PAVOQUE} \\ (German) \\(5 classes)}}
    &  WavLM     & 95.04 & \textbf{98.81} & +3.97   & 92.53 & \textbf{98.27} & +6.20   & 93.55 & \textbf{98.10} & +4.86  \\
    &  HuBERT    & 91.27 & \textbf{98.07} & +7.45   & 87.55 & \textbf{96.63} & +10.37  & 87.29 & \textbf{97.17} & +11.32 \\
    &  Wav2vec2  & 73.44 & \textbf{98.07} & +33.54  & 59.35 & \textbf{96.80} & +63.10  & 56.88 & \textbf{96.97} & +70.48 \\
    &  Emotion2Vec & 93.93 & \textbf{98.44} & +4.80  & 90.77 & \textbf{97.26} & +7.15   & 91.35 & \textbf{97.59} & +6.83  \\
    &  BEATs     & 98.71 & \textbf{98.90} & +0.19   & 97.85 & \textbf{97.99} & +0.14   & 98.05 & \textbf{98.18} & +0.13  \\
    &  CLAP      & 94.21 & \textbf{98.16} & +4.19   & 90.44 & \textbf{96.75} & +6.98   & 91.00 & \textbf{97.25} & +6.87  \\
    \midrule
    \multirow{6}{*}{\shortstack{{CaFE} \\ (French) \\(7 classes)}}
    &  WavLM     & 52.41 & \textbf{77.01} & +46.94  & 54.44 & \textbf{79.12} & +45.33  & 52.65 & \textbf{76.71} & +45.70 \\
    &  HuBERT    & 46.52 & \textbf{47.59} & +2.30   & 45.85 & \textbf{48.38} & +5.52   & 44.32 & \textbf{47.12} & +6.32  \\
    &  Wav2Vec2  & 32.09 & \textbf{35.83} & +11.65  & 37.97 & \textbf{35.45} & -6.64   & 31.80 & \textbf{35.86} & +12.77 \\
    &  Emotion2Vec & 46.52 & \textbf{71.66} & +54.04 & 47.09 & \textbf{70.96} & +50.69  & 44.67 & \textbf{71.25} & +59.50 \\
    &  BEATs     & 72.73 & \textbf{82.89} & +13.97  & 72.57 & \textbf{81.89} & +12.84  & 71.22 & \textbf{81.74} & +14.77 \\
    &  CLAP      & 51.34 & \textbf{59.36} & +15.62  & 55.40 & \textbf{58.70} & +5.96   & 50.85 & \textbf{58.71} & +15.46 \\
    \midrule
    \bottomrule
    \end{tabular}
    \end{adjustbox}
    \end{threeparttable}
\end{table*}

\subsection{Experimental Setup}

\paragraph{Datasets}
    We conduct a extensive evaluation on 14 public speech emotion datasets\footnote{\url{https://huggingface.co/datasets/amu-cai/CAMEO}, \par
\url{https://github.com/numediart/EmoV-DB}, \par\url{https://huggingface.co/datasets/EdwardLin2023/AESDD}} comprising 5 for English and 9 for non-English languages. The diversity of these datasets enables a thorough assessment of the effectiveness and robustness of the proposed method, as well as its language independence grounded in physiological analysis. All datasets are established benchmarks and consist of acted and simulated recordings, consistent with standard practices in the SER research (See Appendix A for details about the datasets).

\paragraph{Baselines}
    To rigorously quantify the additive contribution of our proposed modules, we establish a primary baseline, Backbone-only. This baseline precisely mirrors the backbone branch of our framework: it uses the identical frozen state-of-the-art SSL backbones, the same masked attention pooling layer, and the final classifier. As detailed in Appendix B, we select six representative SSL backbones covering four distinct pre-training paradigms:
    \begin{itemize}[noitemsep, topsep=0pt]
    \item Speech-centric: Wav2vec2\cite{baevski2020wav2vec}, HuBERT\cite{10.1109/TASLP.2021.3122291}, WavLM\cite{chen2022wavlm}.
    \item Emotion-specific: Emotion2Vec \cite{ma-etal-2024-emotion2vec}.
    \item General audio: BEATs \cite{10.5555/3618408.3618611}.
    \item Cross-modal: CLAP (audio encoder only) \cite{elizalde2023clap}.
    \end{itemize}
    Although our task is audio only SER, we include encoders from general audio and audio–text paradigms in addition to speech and emotion models. This design enables a comprehensive stress test that examines whether representations learned from diverse acoustic scenes or language grounding capture emotion features differently from speech models, and it verifies that VAP-informed features supply a complementary representation grounded in the physical structure of the signal under emotional change and in physiological modulations, improving backbones regardless of pre-training domain.

\paragraph{Evaluation Metrics}
    To provide a extensive and robust evaluation of model performance, we report WA, UA, and F1. WA measures overall correctness and is dominated by majority classes, while UA and F1 give equal weight to each emotion class by averaging per-class recall and F1, respectively. All metrics are computed from the confusion matrix $\mathcal{M}$ following standard definitions (see Appendix C for details).

\paragraph{Implementation Details}

    All models were implemented in PyTorch and trained on a single NVIDIA H100 GPU. We utilize a two-stage training pipeline with the AdamW optimizer. The SSL backbones remain frozen as latent feature extractors throughout all experiments. Data is partitioned into training, validation, and testing sets using a $7:1:2$ ratio, with results averaged over five runs with different random seeds. Detailed implementation settings are given in Appendix~D.

\subsection{Main Results and Analysis} 

    We present a extensive validation of our framework in \Cref{tab:multi_ds_ssl} (English datasets) and \Cref{tab:multi_ds_ssl_en} (non-English datasets), covering 14 datasets, 10 languages, and 6 distinct SSL backbones. The results show that our model consistently and substantially outperforms the Backbone-only baselines. These improvements indicate that the VAP-informed vocal feature representation provides a rich, complementary source of emotional features that enhances the emotion representations of existing large scale SSL models. In particular, \ua{} and \mf{} rise in concert with \wa{}, indicating that the benefits are not confined to dominant classes but reflect more balanced performance per class. Moreover, the model yields large relative improvements for weaker backbones (e.g., +136.07\%~\wa{} on JLCorpus with Wav2vec2) while still providing consistent, positive headroom in high accuracy regimes (e.g., +0.52\%~\ua{} on EmoV-DB with BEATs), which demonstrates its robustness.

    This performance pattern reflects principled feature complementarity and backbone choice. The relative improvement ($\Delta$) is inversely correlated with the baseline’s inherent emotion representation quality. For content centric models (e.g., Wav2vec2, HuBERT), which lack emotion features and often treat emotion modulation as noise, vocal feature captures this signal explicitly, yielding the largest gains (e.g., +54.76\%~\wa{} on EMNS with HuBERT). More importantly, our method also improves emotion specialized backbones such as Emotion2Vec (e.g., +42.85\%~\wa{} on AESDD), indicating that vocal feature supplies complementary emotion features underrepresented in their feature spaces. By injecting other physiological features ($f_{\mathrm{inst}}, \tau_{\mathrm{g}}$), by which such pipelines may underutilize, and aligning representations via CPA, the framework contributes an additional, discriminative signal across backbones.

    The complementarity originates from the language agnostic, physiologically grounded nature of our feature quartet. Improvements are systematically larger on non-English datasets (e.g., +109.17\% \ua{} on nEMO and +84.08\% \ua{} on MESD) because predominantly English-trained SSL backbones suffer from linguistic domain shift, whereas vocal feature representations supply a stable signal rooted in human phonation. This physiological sensitivity also helps disambiguate acoustically confusable emotions. On datasets like RAVDESS, baseline models struggle to separate classes with similar magnitude profiles, but our model responds to underlying production differences (e.g., laryngeal tension) encoded in the fine-grained phase features of $f_{\mathrm{inst}}$ and $\tau_{\mathrm{g}}$. This targeted disambiguation, evidenced by the gain of +39.59\% \ua{} for Emotion2Vec on RAVDESS (58.10\% to 81.10\%) shows that our physiological approach directly addresses the known weaknesses of magnitude centric SER. For short or lexically sparse materials (e.g., MESD), first-order AM-FM structure and spectral dynamics become more diagnostic, yielding large lifts for speech centric SSL (+84.08\% \ua{} on Wav2vec2).

\subsection{Ablation Study}

    To rigorously validate our design principles and the efficacy of each component, we conduct extensive ablation experiments on CREMA-D. We adopt WavLM as the representative SSL backbone due to its robust content and paralinguistic representations. We designed diverse variants to comprehensively test the model: 1) Baseline Comparisons: \textbf{WavLM Only} is a frozen backbone baseline where only the downstream classifier is trained; \textbf{Fine-tuned WavLM} employs full end-to-end backbone optimization without physiological modules; and Normal CNN replaces the quaternion-valued QSE with a standard real-valued CNN of comparable parameters to test the necessity of quaternion algebra. 2) PhysioSER Configurations: \textbf{PhysioSER} is our proposed lightweight framework combining a frozen WavLM; while \textbf{PhysioSER-Full} is an end-to-end variant where the entire architecture is optimized jointly. 3) Component Ablations: feature quartet variants remove specific components ($f_{\mathrm{inst}}, \tau_{\mathrm{g}}$, $M, \rho$) to test feature integrity; and architectural variants evaluate pooling strategies, normalization schemes, and the CPA module's contribution.

    We analyze the performance implications below:

    \textbf{1) Parameter Efficiency:} We first compare the efficiency of PhysioSER with the Fine-tuned WavLM. As shown in \Cref{tab:main_compact}, Fine-tuned WavLM requires 94.61M parameters to attain 73.66\% WA. In contrast, PhysioSER achieves a superior 75.20\% WA with only 2.03M trainable parameters (2.1\% of the baseline), demonstrating a far more efficient form of knowledge injection. Furthermore, PhysioSER-Full yields the peak performance of 76.41\% WA, which confirms that vocal feature contributes complementary emotion features, including the phase gradient information in $f_{\mathrm{inst}}$ and $\tau_{\mathrm{g}}$, that standard SSL fine-tuning does not fully capture.
    
    \textbf{2) Necessity of Quaternion Algebra:} To validate the effectiveness of the QSE, we compare PhysioSER with Normal CNN (\Cref{tab:ablation_cremad}). While Normal CNN improves over WavLM Only, it trails our quaternion model by 3.22\% WA. This gap confirms the efficacy of the Hamilton product, which induces structured parameter sharing and principled mixing of the physiological features ($M$, $\rho$, $f_{\mathrm{inst}}$, $\tau_{\mathrm{g}}$) more effectively than standard real-valued CNN.

    \textbf{3) Integrity of the Feature Quartet:} Evaluations of Feature Subsets reveal that removing phase information (Magnitude-only) causes a minor 0.80\% drop, whereas discarding magnitude (Phase-only) leads to a substantial 6.25\% decline. While $M$ and $\rho$ are dominant, the drop in Magnitude-only proves the unique complementary value of phase gradients ($f_{\mathrm{inst}}, \tau_{\mathrm{g}}$). Notably, removing any single component consistently degrades performance (e.g., -2.95\% w/o $\tau_{\mathrm{g}}$), validating the quartet as a minimal yet comprehensive physiologically grounded set.

    \textbf{4) Robustness of Architectural Choices:} Finally, we examine the architectural variants. Replacing masked attention pooling with mean or max pooling reduces WA by 2.86\% and 12.33\% respectively, confirming the need for adaptive weighting. Similarly, standard or shared batch normalization underperforms our independent-channel QBN. Regarding alignment, removing CPA pre-training reduces WA by 1.97\%, while end-to-end alignment training degrades it by 2.67\%, which validates that our two-stage protocol yields a more reliable shared embedding space.


\begin{table}[t]
    \centering
    \caption{Results on CREMA-D with parameter efficiency.}
    \label{tab:main_compact}
    \begin{adjustbox}{width=\columnwidth}
    \begin{tabular}{lcccc}
    \toprule
    Method & WA (\%) $\uparrow$ & UA (\%) $\uparrow$ & F1 (\%) $\uparrow$ & Params \\
    \midrule
    WavLM Only (Frozen + Classifier)          & 69.69 & 70.22 & 69.79 & 0.23M \\
    \textbf{PhysioSER (Frozen WavLM)}             & \textbf{75.20} & \textbf{75.62} & \textbf{75.63} & \textbf{2.03M} \\
    \textit{$\Delta$ (relative, \%)}          & \textit{+7.91} & \textit{+7.69} & \textit{+8.37} & \textit{--} \\
    \midrule
    Fine-tuned WavLM                          & 73.66 & 74.07 & 73.87 & 94.61M \\
    \textbf{PhysioSER-Full }      & \textbf{76.41} & \textbf{76.87} & \textbf{76.55} & \textbf{96.64M} \\
    \textit{$\Delta$ (relative, \%)}           & \textit{+3.73} & \textit{+3.78} & \textit{+3.63} & \textit{--} \\
    \bottomrule
    \end{tabular}
    \end{adjustbox}
\end{table}

\begin{table}[t]
    \centering
    \caption{Ablation results on CREMA-D.}
    \label{tab:ablation_cremad}
    \setlength{\tabcolsep}{4.5pt}
    \renewcommand{\arraystretch}{0.8}
    \scriptsize
    \begin{threeparttable}
    \begin{adjustbox}{width=\columnwidth}
    \begin{tabular}{l|cc|cc|cc}
    \toprule
    Method 
    & \wa~(\%) $\uparrow$ & $\Delta$
    & \ua~(\%) $\uparrow$ & $\Delta$
    & \mf~(\%) $\uparrow$ & $\Delta$ \\
    \midrule
    \textbf{PhysioSER} 
    & \textbf{75.20} & --
    & \textbf{75.62} & --
    & \textbf{75.63} & -- \\
    Normal CNN
    & 72.78 & -3.22 
    & 73.04 & -3.41 
    & 73.06 & -3.40 \\
    \midrule
    \multicolumn{7}{c}{\textit{Quaternion feature quartet}} \\
    \addlinespace[2pt]
    Magnitude-only                             & 74.60 & -0.80 & 74.98 & -0.84 & 74.38 & -1.65 \\
    Phase-only                                 & 70.50 & -6.25 & 70.99 & -6.12 & 70.68 & -6.55 \\
    w/o $M$                                    & 73.12 & -2.77 & 73.57 & -2.71 & 73.30 & -3.08 \\
    w/o $\rho$                                 & 74.26 & -1.25 & 74.44 & -1.56 & 74.37 & -1.67 \\
    w/o $f_{\mathrm{inst}}$                    & 74.26 & -1.25 & 74.48 & -1.51 & 74.27 & -1.80 \\
    w/o $\tau_{\mathrm{g}}$                    & 72.98 & -2.95 & 73.15 & -3.27 & 72.88 & -3.64 \\
    \midrule
    \multicolumn{7}{c}{\textit{Sentence-level aggregation}} \\
    \addlinespace[2pt]
    Mean pooling                               & 73.05 & -2.86 & 73.18 & -3.23 & 73.58 & -2.71 \\
    Max pooling                                & 65.93 & -12.33 & 66.10 & -12.59 & 66.53 & -12.03 \\
    \midrule
    \multicolumn{7}{c}{\textit{Normalization}} \\
    \addlinespace[2pt]
    Standard BN                                & 73.86 & -1.78 & 74.31 & -1.73 & 74.02 & -2.13 \\
    Shared BN                                  & 73.79 & -1.88 & 74.27 & -1.78 & 73.69 & -2.57 \\
    \midrule
    \multicolumn{7}{c}{\textit{Alignment (CPA) and training regime}} \\
    \addlinespace[2pt]
    End-to-end (w/ CPA)                       & 73.19 & -2.67 & 73.54 & -2.75 & 73.50 & -2.82 \\
    w/o CPA                                   & 73.72 & -1.97 & 74.18 & -1.90 & 73.75 & -2.49 \\
    \bottomrule
    \end{tabular}
    \end{adjustbox}
    \end{threeparttable}
\end{table}

\begin{figure}[t]
    \centering
    \begin{subfigure}[b]{0.24\textwidth}
        \centering
        \includegraphics[width=\textwidth]{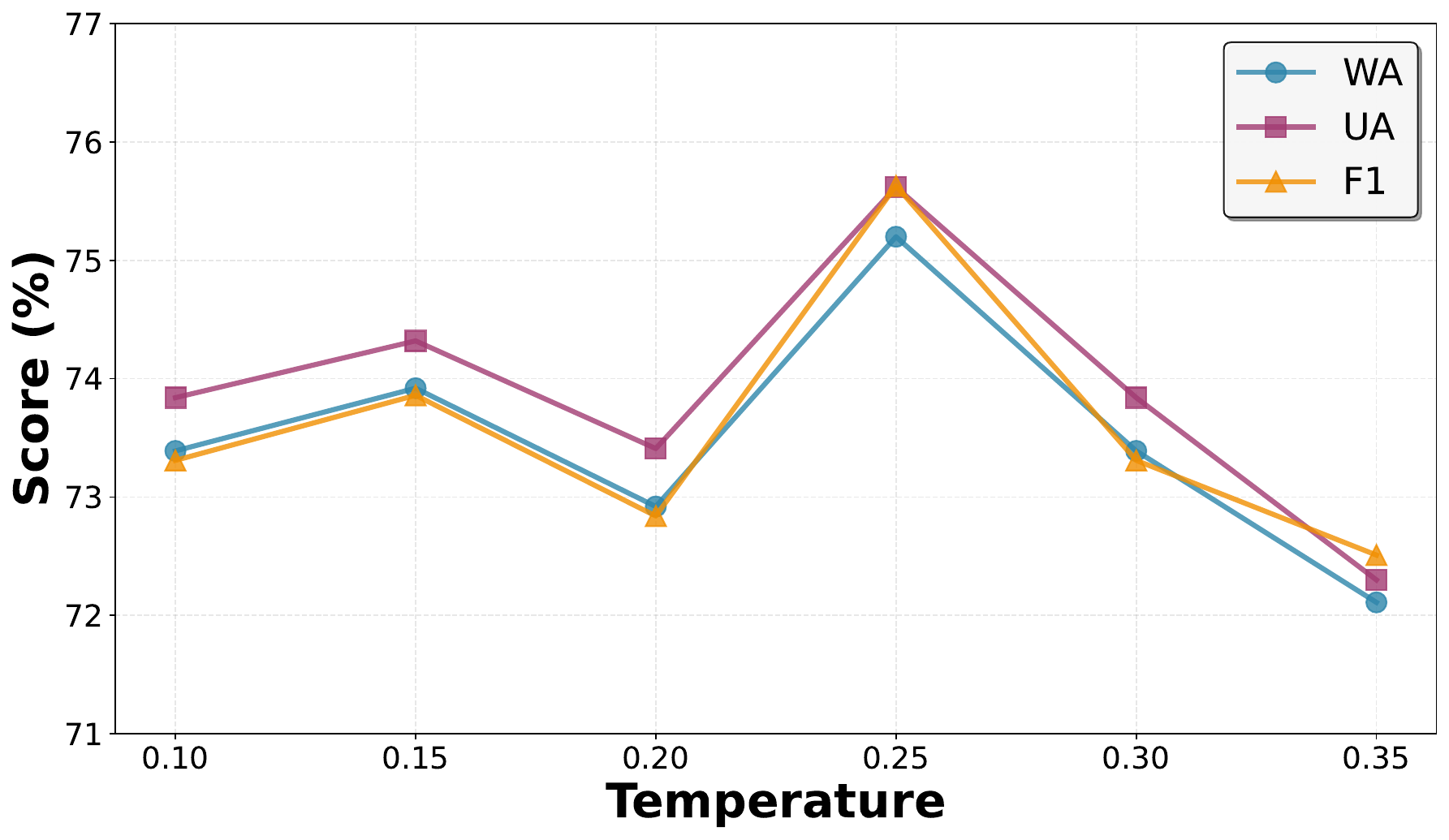}
        \caption{CPA Temperature ($\eta$)}
        \label{fig:hyperparams-temp}
    \end{subfigure}
    \hfill
    \begin{subfigure}[b]{0.24\textwidth}
        \centering
        \includegraphics[width=\textwidth]{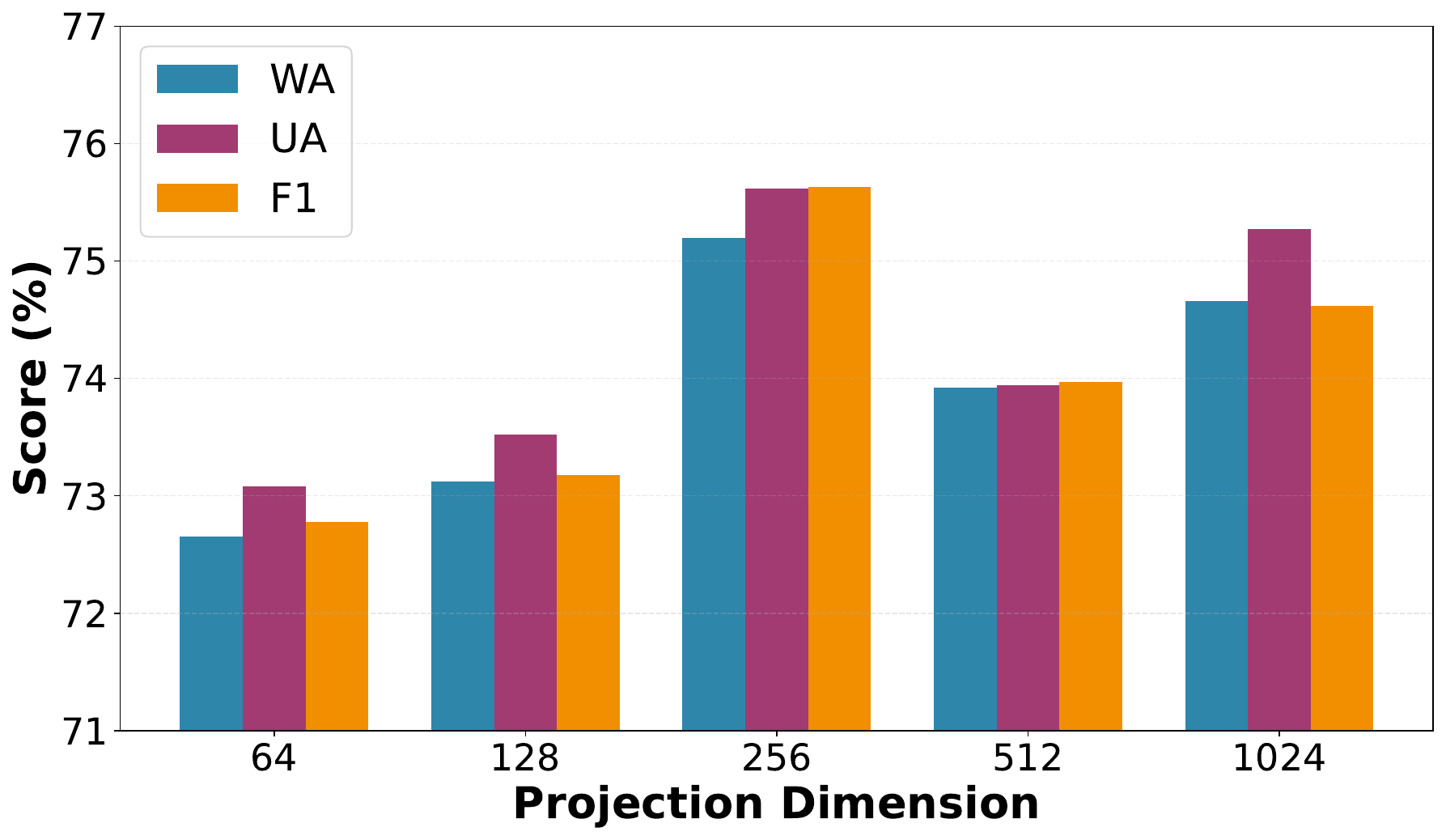}
        \caption{CPA Alignment Dimension}
        \label{fig:hyperparams-dim}
    \end{subfigure}
    \hfill
    \begin{subfigure}[b]{0.48\textwidth}
        \centering
        \includegraphics[width=\textwidth]{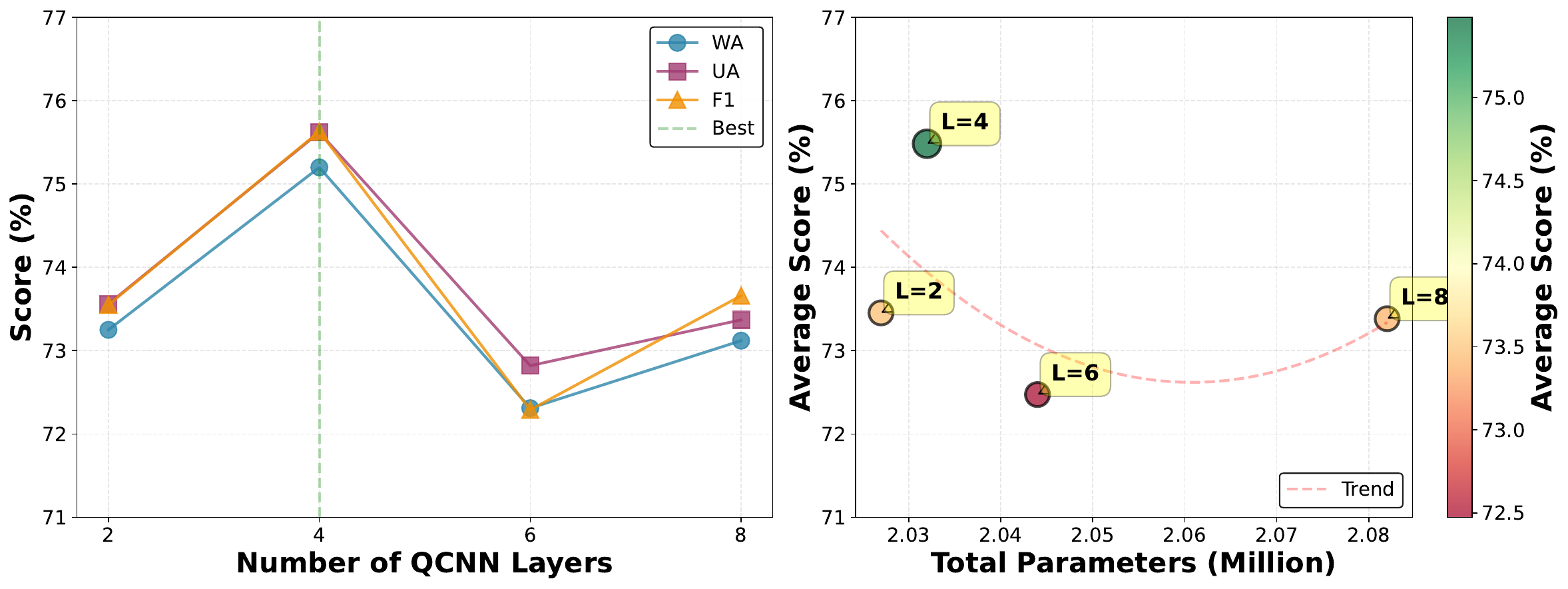}
        \caption{QSE Depth (L)}
        \label{fig:hyperparams-layers}
    \end{subfigure}
    \caption{Hyperparameter sensitivity analysis on CREMA-D. (a) Performances on different temperature $\eta$. (b) Performances on different CPA alignment dimension $d_{\mathrm{align}}$. (c) Performances and complexity on different QSE depth $L$.}
    \label{fig:hyperparams}
\end{figure}

\subsection{Hyperparameter Sensitivity Analysis}

    As shown in \Cref{fig:hyperparams}, we analyze the sensitivity of key hyperparameters within our framework, specifically the CPAF temperature $\eta$, the CPA alignment dimension $d_{\mathrm{align}}$, and the QSE depth $L$, with results reported on CREMA-D.

    \textbf{CPA Temperature ($\eta$):} In the InfoNCE loss, $\eta$ rescales cosine similarities before the softmax and thus sets the effective angular margin between positive and negative pairs. \Cref{fig:hyperparams-temp} shows a clear optimum at $\eta\!=\!0.25$. Lower $\eta$ sharpens the softmax, overweights hard negatives, increases gradient variance, and makes the contrastive learning sensitive to outliers and false negatives. Higher $\eta$ flattens the distribution, weakens gradients, compresses inter-utterance margins, and fails to separate positives from their nearest negatives. Both degrade performance relative to $\eta\!=\!0.25$.

    \textbf{CPA Alignment Dimension ($d_{\mathrm{align}}$):} \Cref{fig:hyperparams-dim} analyzes the dimension of the shared projection space. Performance peaks at $d_{\mathrm{align}}\!=\!256$, which outperforms the other settings. Lower dimensions impose an overly tight bottleneck and discard discriminative features. Higher dimensions introduce excess capacity, weaken the contrastive pressure, and reduce cross-view alignment, yielding smaller margins. Thus, $d_{\mathrm{align}}\!=\!256$ provides a compact space that preserves information while enforcing effective alignment.

    \textbf{QSE Depth ($L$):} \Cref{fig:hyperparams-layers} evaluates depth against accuracy and complexity. We get the best performance at $L=4$ and declines for deeper models ($L\!=\!6$ and $L\!=\!8$), which suggests overfitting. The right panel shows that $L\!=\!4$ lies on a favorable point of the performance–parameter curve, achieving the highest WA with a modest parameter count (2.032M). Increasing depth to $L\!=\!8$ raises the parameters to 2.082M yet reduces accuracy.

\begin{figure}[t]
    \centering
    \includegraphics[width=0.45\textwidth]{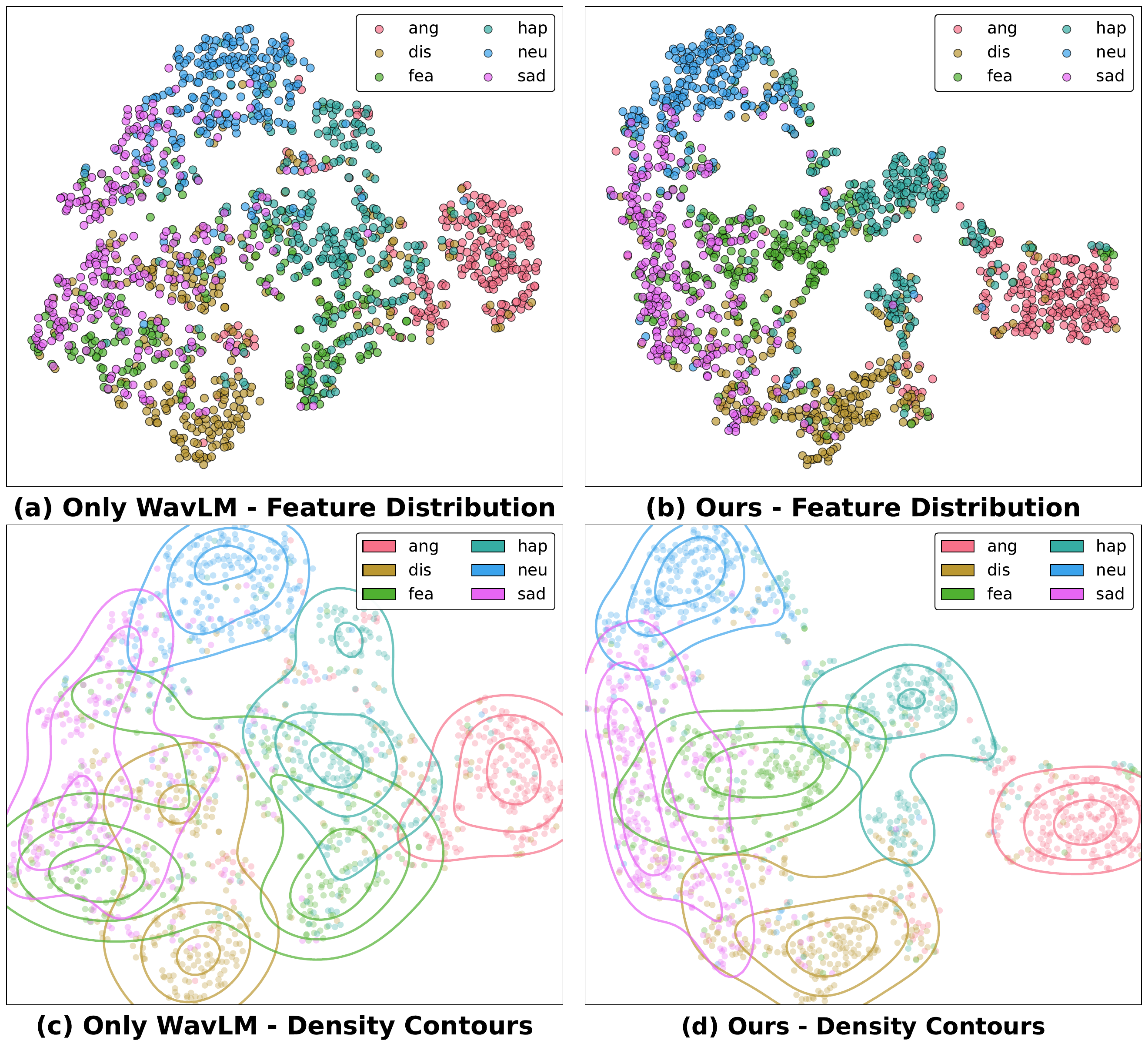}
    \caption{Visualization of test set feature clusters on CREMA-D.}
    \label{fig:visualization-cluster}
\end{figure}

\subsection{Feature Visualization and Physiological Discriminability}

    To validate the discriminative structure of the learned embeddings, we project the test set features from WavLM-Only baseline and our model into two dimensions using t-SNE shown in \Cref{fig:visualization-cluster}. The baseline embeddings in \Cref{fig:visualization-cluster}(a) and (c) appear diffused with substantial inter-class overlap and poorly defined boundaries. In contrast, our embeddings in \Cref{fig:visualization-cluster}(b) and (d) form tighter clusters with clearer inter-class separation. Quantitative clustering indices confirm this improvement: the Silhouette score increases from 0.096 to 0.198, the Davies–Bouldin index improves from 3.77 to 1.44, and the Calinski–Harabasz score rises from 387.2 to 577.0.
    
\begin{figure}[t]
    \centering
    \includegraphics[width=0.85\columnwidth]{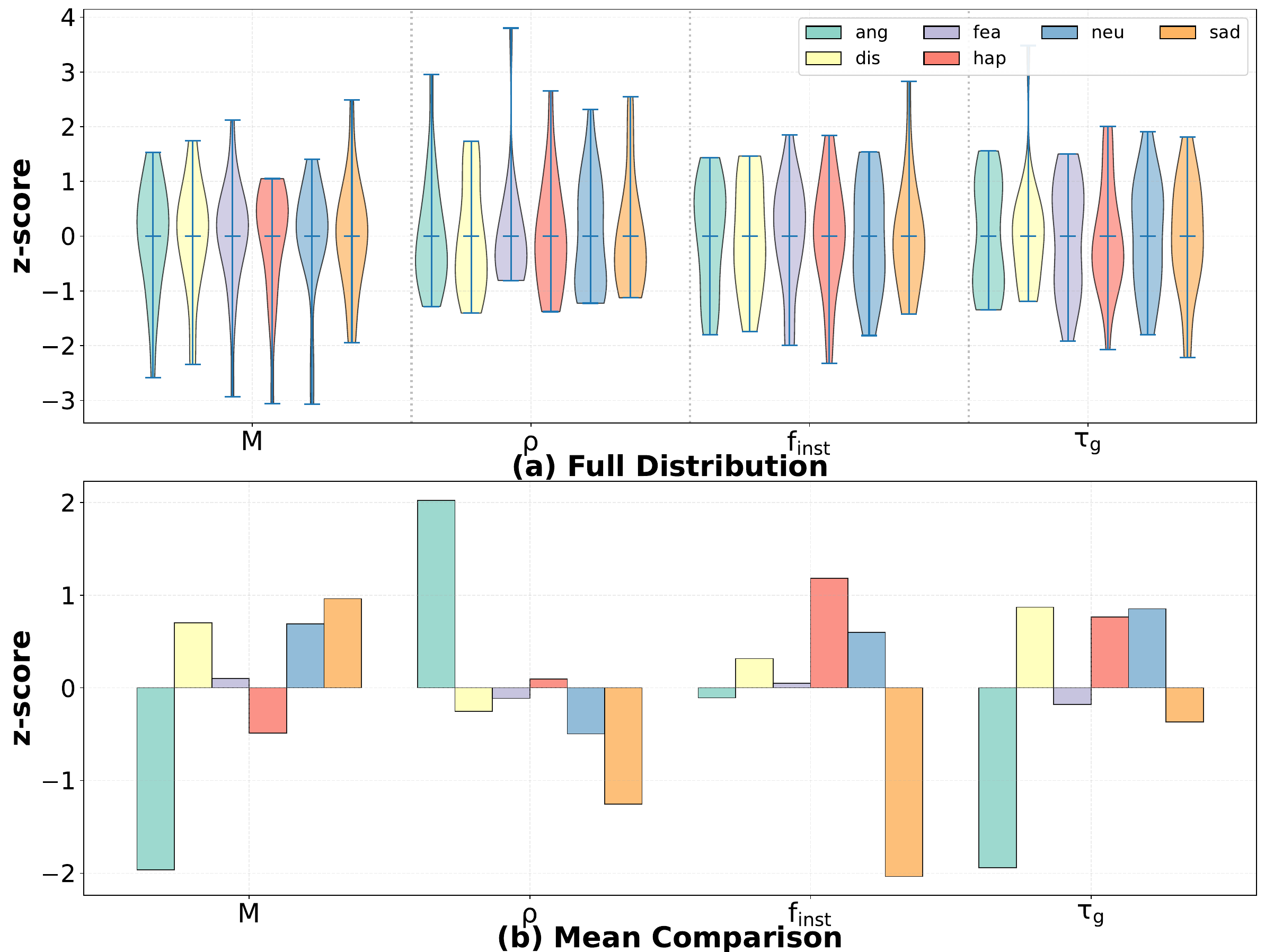}
    \caption{Analysis of the vocal features quartet $\{M, \rho, f_{\mathrm{inst}}, \tau_g\}$ across emotions on the CREMA-D testset.}
    \label{fig:feature-fingerprints}
\end{figure}

    This enhanced separation is driven by the discriminative power of the VAP-informed quartet $\{M,\ \rho,\ f_{\mathrm{inst}},\ \tau_{\mathrm{g}}\}$, whose class-conditional statistics are analyzed in \Cref{fig:feature-fingerprints}. While the full distributions (\Cref{fig:feature-fingerprints}a) exhibit partial overlap, the joint quartet provides sufficient separability where single features fail. For instance, magnitude ($M$) alone is ambiguous for separating disgust, neutral, and sadness. However, the phase-derived and dynamic components resolve these conflicts: anger is distinguished by sharp temporal energy modulation (high $\rho$) and low group delay ($\tau_{\mathrm{g}}$); happiness is characterized by elevated instantaneous frequency ($f_{\mathrm{inst}}$) due to increased laryngeal tension; and sadness is marked by the lowest $f_{\mathrm{inst}}$ and reduced $\rho$, reflecting flatter prosody. Taken together, these physiological trends form the compact, multidimensional basis that enables the distinct clustering observed in \Cref{fig:visualization-cluster}.
    
\subsection{Validation on An Emotional Humanoid Robot}
\label{sec:robot-experiments}
\begin{figure}[t]
    \centering
    \includegraphics[width=\columnwidth]{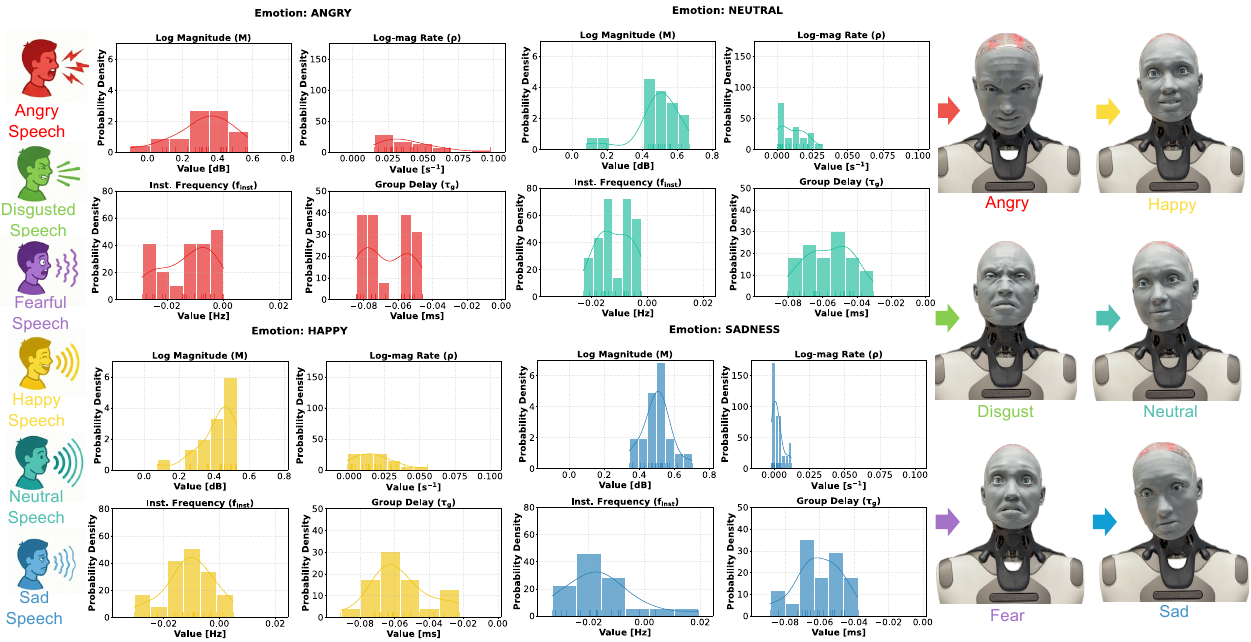}
    \caption{Real-time deployment of PhysioSER on the Ameca. The system maps speech-driven emotional states to corresponding facial expressions.}
    \label{fig:robot-experiments}
\end{figure}

    To evaluate the practical utility, we integrated PhysioSER into the perception-action loop of the humanoid robot Ameca for real-time affective sensing. As demonstrated in our experimental video\footnote{\href{https://drive.google.com/drive/folders/1zs1myYPPKARx4OP9IzeASVtTzQOOF4pl?usp=sharing}{Video showing real-time PhysioSER deployment effect}.} and illustrated in \Cref{fig:robot-experiments}, PhysioSER captures the nuanced vocal emotional features and processes continuous audio streams and maps vocal cues directly to Ameca's facial emotions without relying on automatic speech recognition or linguistic analysis.
    
    Enabled by PhysioSER, Ameca can recognize vocal emotional cues with negligible latency, further transferred to synchronized facial expressions congruent with the recognized vocal affective states. This validation confirms the computational efficiency of the light-weighted PhysioSER framework, demonstrating its ability to maintain stable real-time operations while preserving the accuracy gains observed in offline benchmarks. PhysioSER offers a practical and interpretable vocal emotion tool for developing multimodal humanoid emotions \cite{Cao2025HumanoidAI}.

\section{Conclusion}
\label{sec:conclusion}
    We present PhysioSER to enhance SER informed by voice anatomy and physiology (VAP) integrating vocal feature decomposition and general latent speech representations from a frozen, pretrained semi-supervised learning (SSL) model. Specifically, we formalize a compact, physiologically and physically interpretable feature quartet spanning vocal-tract timbre, temporal energy modulation, laryngeal oscillation rate, and phase dispersion, and embed it by a quaternion spectrotemporal encoder, where Hamilton-structured convolutions preserve structured cross-component coupling among these complementary cues. In parallel, an SSL workflow enables general-purpose latent representations, while the Contrastive Projection and Alignment framework aligns the two utterance-level embeddings without erasing their complementary inductive biases. These parallel representations are fused by a parameter efficient mechanism with a classification head for SER. Extensive experiments across 14 datasets, 10 languages, and 6 backbones show consistent improvements over backbone-only systems. Moreover, gains on non-English datasets indicate language independent robustness. Ablation studies verify the contribution of each module and show that physiologically derived signals from emotional vocal behaviors encode emotion while remaining interpretable, linking vocal changes to underlying phonatory and articulatory mechanisms. Furthermore, the successful deployment on the humanoid Ameca demonstrates the practical efficiency and effectiveness of PhysioSER for real-time affective sensing. A current limitation is the reliance on short-time Fourier transform windowing and resolution choices. Future work will also integrate VAP-informed features into large voice models and extend PhysioSER to  additional physiological signals for other speech tasks for humanoids.

\ifCLASSOPTIONcaptionsoff
  \newpage
\fi

\bibliographystyle{IEEEtran}
\bibliography{IEEEabrv,Bibliography}

\end{document}